\documentclass[fleqn,usenatbib]{mnras}

\usepackage{newtxtext,newtxmath}

\usepackage[T1]{fontenc}

\DeclareRobustCommand{\VAN}[3]{#2}
\let\VANthebibliography\thebibliography
\def\thebibliography{\DeclareRobustCommand{\VAN}[3]{##3}\VANthebibliography}

\usepackage{graphicx}	
\usepackage{amsmath}	
\usepackage{relsize}
\usepackage{booktabs}
\usepackage{xcolor}

\title[Magneto-thermo-elastic mountains]{Numerical calculations of neutron star mountains supported by crustal lattice pressure}

\author[T. J. Hutchins \& D. I. Jones]{
T. J. Hutchins,$^{1}$\thanks{E-mail: T.Hutchins@soton.ac.uk}
D. I. Jones,$^{1}$
\\
$^{1}$Mathematical Sciences and STAG Research Centre, University of Southampton, Southampton SO17 1BJ, United Kingdom
}

\date{Accepted XXX. Received YYY; in original form ZZZ}

\pubyear{\the\year{}}

\begin{document}
\label{firstpage}
\pagerange{\pageref{firstpage}--\pageref{lastpage}}
\maketitle

\begin{abstract}
Gravitational waves may set the spin frequencies of neutron stars in low-mass X-ray binaries (LMXBs). One mechanism for facilitating such emission is the formation of a mass asymmetry - or `mountain' - supported by elastic strains driven by thermal gradients. Most studies have focused either on the origin of the elastic strains or the temperature asymmetry in isolation, and have not considered the entire formation process. In previous work, we showed that anisotropic heat transport in magnetised accreting neutron stars can source a significant temperature asymmetry, and made rough estimates that suggested temperature-induced perturbations in the pressure supplied by the crustal lattice may be competitive with the widely known model of temperature-induced capture-layer shifts. 
In this paper we carry out detailed calculations to properly explore this scenario.  We self-consistently calculate the temperature asymmetries, the perturbations in crustal lattice pressure, and the mass asymmetries within a single framework. For the first time, we make use of the set of realistic equations of state from the Brussels–Montreal nuclear energy-density functionals BSk19, BSk20, and BSk21 which describe all regions of accreting neutron stars in a thermodynamically consistent, unified way. We find these mountains are too small to dictate the spin-equilibrium of LMXBs, and estimate the level of gravitational wave emission they produce.

\end{abstract}




\section{Introduction}

With each passing year, the catalog of gravitational wave (GW) observations from the inspiral and merger of compact binary systems continues to grow \citep{Abac_2026_GWTC4}. And with it, continuing advancements in the sensitivity of LIGO-Virgo-KAGRA instruments have also led to much anticipation for the first detection of quasi-monochromatic \textit{continuous} gravitational waves (CGWs).

Rapidly spinning neutron stars harboring a quadrupolar deformation are one such source of CGWs: see \citet{Jones_Riles_2025} for a review. The deformations are typically characterized either as `magnetic' or `elastic' mountains; where a mass asymmetry is supported either by strong magnetic fields, or by elastic strains that build up in the star's solid crust, respectively.

In assessing the likelihood of detecting CGWs from an elastic mountain in particular, there are two important considerations one needs to make. Given the crust will have a finite shear strength, the first of these is rather simple: what exactly is the largest mountain the crust could feasibly maintain before cracking?

Whether such a `maximum mountain' could be detectable has been the subject of a number of studies \citep{Ushomirsky_2000, Owen_2005, Haskell_2006, Johnson_2013, Gittins_2021, Gittins_2021_relativity, Morales_2022}, and the general consensus places the maximum ellipticity (the fractional difference in the moments of inertia of the star) at around $\varepsilon_{\text{max}} \sim 10^{-7}$. In principle, this is large enough for (elastic) mountains to be seen in current-generation LIGO instruments. This value however is really more of an upper bound, set by the crustal strength (see e.g. \citealt{Horowitz_2009}). Most notably, maximum mountain calculations \textit{do not} address the underlying physical mechanisms responsible for generating the mass asymmetry in the first place.

The question of what ``realistic'' elastic mountain sizes might look like therefore remains uncertain, and leads naturally to the second important question: what specific physical processes taking place inside the star are capable of sourcing the mountain in the first place? Only a small number of detailed analyses of `likely' elastic mountain sizes can be found in the literature. Those that do have tended to focus on exploring the physical origins behind a specific subtype of elastic mountain referred to as a `thermal mountain', where elastic strains are sourced from large-scale temperature gradients \citep{Bildsten_1998}.

Such studies have typically focused either on a particular mechanism to induce the elastic strains (e.g. \citealt{Ushomirsky_2000, Jones_2025}), or on the origin of the temperature asymmetry (e.g. \citealt{Singh_2020, Osborne_2020, Hutchins_2023}), independently of one another.  What is really required is a fully self-consistent treatment; one and the same stellar model being used to calculate the thermal \textit{and} elastic solutions. This is precisely what we do in this paper. The only other existing such treatment is that of \citealt{Li_2025}, who looked at mountain formation in ultra-luminous X-ray (ULX) pulsars. 

We differ from all previous analyses of thermal mountain formation in one important regard. Previous works have, in some way or another, relied upon the `wavy capture layer' scheme developed by \citet{Ushomirsky_2000} (hereafter the `UCB model'; see Sec. \ref{SubSec: Origin of the mass quadrupole moment}) to source the elastic strains, whereby electron capture reactions take place at lower density in regions of the crust that are locally hotter. As first pointed out in \citet{Jones_2025} however, recent calculations of the crustal equation of state (EOS) predict far fewer electron capture events in accreting neutron stars, casting significant doubts on the UCB model producing mass quadrupoles as large as initially predicted. 

In \citet{Jones_2025} we therefore proposed an alternative scheme for generating the required mass asymmetry altogether. Rather than exploiting EOS-dependent electron capture reactions, we considered a piece of the crustal pressure that is tied to the crystal lattice itself. Like the electron capture events, this `crustal lattice pressure' is also temperature-dependent, but is in fact relatively \textit{insensitive} to the crustal composition. At the time, we made only a rudimentary comparison of the pressure perturbations generated by the UCB model and our own, and performed no explicit calculation of the mass quadrupoles that arise from thermal perturbations in the lattice pressure. 

As such, in this paper we present a new, self-consistent estimate of the size of elastic neutron star (NS) mountains that encompasses the entire formation process, combining methodologies from both \citet{Hutchins_2023} and \citet{Jones_2025}, into the framework originally developed by \citet{Ushomirsky_2000}. In doing so we shall complete the work initially presented in \citealt{Osborne_2020}, 
computing the ellipticity from the full function $\delta T(r)$ that arises from anisotropic heat conduction due to internal magnetic fields. 

We henceforth term the deformations that arise from a combination of magnetically-induced anisotropic heat conduction and the thermal pressure as ‘magneto-thermo-elastic mountains’, rather than simply ‘thermal mountains’, which was the term originally coined in \citet{Bildsten_1998}, but perhaps more precisely, should be termed `thermo-elastic' mountains since they did not ascribe a magnetic field origin for the temperature asymmetry. 

In Section \ref{Sec: Context} we begin with a review of prior efforts on modelling thermo-elastic mountain formation, before going on to outline the steps required to produce a self-consistent calculation. In Section \ref{Sec: Models of Accreting Neutron Stars} we detail the relevant microphysics for our calculation. In Section \ref{Sec: Hydrostatic Structure} we then fix the hydrostatic structure of neutron stars which are assumed to be steadily accreting. 
Models of the background and perturbed thermal structure are described in Sections \ref{Sec: Thermal Structure of Accreting Neutron Stars} and \ref{Sec: Perturbations in the Thermal Structure} respectively.  In Section \ref{Sec: Quadrupolar Deformations of Accreting Neutron Stars} we then calculate the mass quadrupole moment associated with an elastic readjustment of the crust in response to the aforementioned temperature perturbations produced by the magnetic field coupled to the thermal component of the lattice pressure. We present results for the ellipticity as a function of the rate of mass accretion and amount of shallow crustal heating, which warms the crust. We then compare sizes of mountains calculated here (i.e. via the lattice pressure) with those calculated in \citet{Ushomirsky_2000} (i.e. via capture layer shifts). In Section \ref{Sec: Model Improvements} we describe a few areas where our model might be improved, before summarizing in Section \ref{Sec: Summary}.

\section{Sources of asymmetry}
\label{Sec: Context}

A perturbation in the density profile of an (otherwise) spherically-symmetric neutron star will lead it to develop a set of multipole moments defined by \citep{Ushomirsky_2000}

\begin{equation}\label{eq: Mass Quadrupole}
    Q_{\ell m} = \int \delta \rho_{\ell m} (r) \, r^{l+2} \, \text{d}r \, , 
\end{equation}

\noindent
where ($\ell, m$) defines the harmonic mode of the perturbation. 

The specific scenario relevant to the present work is that of a neutron star that is rigidly rotating, meaning we are interested primarily in the case where $\ell = m = 2$ (assuming rotation about the $z$-axis), since this will give rise to quadrupolar gravitational radiation - the strongest, and lowest-order component.

The mass quadrupole however is not necessarily a particularly intuitive measure of the size of the mountain. Instead, mountain sizes are often quantified, as above, in terms of the ellipticity, which is related to the magnitude of $Q_{22}$ as

\begin{equation}\label{eq: Ellipticity}
    \varepsilon = \frac{(\delta I_{xx} - \delta I_{yy})}{I_{zz}} = \sqrt{\frac{8\pi}{15}}\frac{Q_{22}}{I_{zz}} \, ,
\end{equation}

\noindent
where $\delta I_{ij} \equiv  \int \delta \rho(r, \theta, \phi) \, r_i r_j \, d^3 V$, and $I_{zz}$ is the principal stellar moment of inertia. In what follows, we shall assume (for simplicity) that $I_{zz} \sim 10^{45}$ g cm$^{2}$. Keep in mind however, that for a real neutron star this value may vary by a factor of a few, depending on its mass and the equation of state (see e.g. Fig. 4 of \citealt{Fantina_2022} and surrounding text). 

A steadily rotating star with such a mass quadrupole emits a continuous gravitational wave signal at twice the spin frequency, with an amplitude linear in $\varepsilon$ and quadratic in the spin frequency; see e.g.\ \citet{st_83}.  Many searches have been carried out for such signals, so far without a detection; see e.g.\  \citet{Abbott_2004, Abbott_2007,Abbott_2017_kpadd, Abbott_2017_kp, Abbott_2017_tensorial_pulsars,Abbott_2019_band,Abbott_2020_Constraint,Abbott_2022_narrow_transient}.

Given the quadratic scaling of the gravitational wave amplitude with spin frequency, millisecond pulsars and neutron stars in 
low-mass X-ray binaries (LMXBs) are promising candidates.  The two populations are believed to be closely related.  LMXBs systems contain a neutron star closely orbiting with a less-compact donor star, from which it accretes matter. A ‘recycling’ scenario posits angular momentum from an accretion disk is transferred to the NS, leading to spin-up torques \citep{Ghosh_1977, Ghosh_1978, Ghosh_1979_a, Ghosh_1979_b}. In principle, extended accretion should have no difficulty spinning up these stars to that of the centrifugal break-up frequency ($\sim$ 1-2 kHz; \citealt{Lattimer_2007}). Yet, the fastest-observed pulsar to-date, PSR J1748-2446ad, rotates at just 716 Hz \citep{Hessels_2006}. Such observations are suggestive of some additional braking torques, preventing the stars from reaching the break-up limit. 

If one assumes that $100\%$ of the implied spin-down energy from the NS is radiated away as CGWs, the required ellipticity for torque balance can be estimated as \citep{Bildsten_1998, Osborne_2020}, 

\begin{equation}\label{eq: Torque Balance Ellipticity}
 \varepsilon_{\text{TB}} \approx 5 \times 10^{-8} \, \bigg( \frac{\Dot{M}}{10^{-9} M_{\odot} \text{yr}^{-1}} \biggr)^{1/2} \, \bigg( \frac{300 \text{Hz}}{\nu_s} \biggr)^{5/2} \, .
\end{equation}

To give some context, a recent search for CGWs directly targeting 20 accreting millisecond pulsars constrained the upper limit of the ellipticity of the pulsar IGR J00291 + 5934 to be $\varepsilon^{95\%} = 3.1 \times 10^{-7}$ \citep{Abbott_2022_AMXP}. Using X-ray flux data taken from \citet{De_Falco_2017} in the years 2004-2015, \cite{Hutchins_2023} estimated the time-averaged mass accretion rate of IGR J00291+5934 to be $1.8 \times 10^{-11}$ M$_{\odot}$ yr$^{-1}$ (assuming the NS has a mass of 1.4 M$_{\odot}$). The neutron star also has an observed spin frequency $\sim 599$ Hz \citep{Sanna_2016}, implying that it would require an ellipticity $ \varepsilon_{\text{TB}} \approx 4 \times 10^{-8}$ in order to balance accretion torques from its $\sim 0.1$ M$_{\odot}$ companion. This is approximately one order of magnitude smaller than the current upper limit imposed by  \citet{Abbott_2022_AMXP}, indicating that current-generation detectors are not yet capable of beating the spin-down limit of this particular pulsar.  

\subsection{Source of the temperature asymmetry}
\label{SubSec: Sources of large-scale temperature asymmetry}

One possible explanation for the origin of the thermal perturbations was described in \citet{Ushomirsky_2000} (hereafter UCB). They argued that lateral variations in the crustal composition of the accreted crust due to asymmetric burning could lead to both asymmetries in heat energy released in the crust, and anisotropies in the crustal thermal conductivity due to varying charge-to-mass ($Z^2/A$) ratios. Whilst they found that temperature asymmetries at the percent level ($\delta T / T \sim 1\%$) could lead to the formation of mass quadrupoles large enough to balance accretion torques under certain circumstances (we will return to this in Section \ref{SubSubSec: Crustal lattice pressure vs. capture layer shifts}), they did not attempt to justify the origin of either the compositional or nuclear heating asymmetries, instead simply assuming their existence \textit{a priori} at the level of 10\%.

More recently, \citet{Singh_2020} took the UCB argument one step further, exploring more rigorously the idea of asymmetric accretion heating. The authors exploited the fact that neutron stars in LMXBs have (albeit weak) external magnetic fields, in order to model the flow of accreted matter onto the polar caps and compute the temperature perturbations that arise due to the build up of accreted matter. Their results focused on the particular pulsar PSR J1023+0038, since an observed increase in its spin-down rate during active episodes of accretion ($\sim 27\%$) has been well documented \citep{Jaodand_2016}. Unfortunately, the mass quadrupoles generated from this mechanism were, in general, too small to explain the spin-down rate of PSR J1023+0038, unless strong shallow heating sources ($\sim 5$ MeV per accreted nucleon) are present in the outermost layers of the crust.

An alternative mechanism was also proposed by \citet{Osborne_2020}, who sought to provide a first-principles calculation for sourcing large-scale temperature variations. They modelled anisotropic heat conduction of relativistic electrons as a result of internal crustal magnetic fields (rather than external fields as in \citealt{Singh_2020}). They concluded that the $1\%$ asymmetry required by UCB to attain torque balance was unlikely to be produced from this mechanism however, finding that asymmetries of the order $10^{-5}\%$ could be produced in the deep crust by a $10^9$ G field.

This work was later extended by \citet{Hutchins_2023}, who exploited the same mechanism as \citet{Osborne_2020}, but allowed for the possibility of the magnetic field to penetrate the core of the star. The presence of a core magnetic field was found to raise the expected level of asymmetry in the deep crust by up to 3 orders of magnitude (but still less than 1\%), by removing the condition that the temperature perturbations must go to zero at the crust-core transition.

\subsection{Source of the mass asymmetry}
\label{SubSec: Origin of the mass quadrupole moment}

Self-consistent estimates of the temperature asymmetry tell only  half the story when it comes to providing physically-motivated estimates of the size of the (thermal) mountain. For such a mountain to form, the thermal asymmetry must induce a pressure imbalance, which in turn leads to a physical displacement of crustal matter. Such displacements can be modelled in terms of a Lagrangian displacement field $\xi^i$, so that one may connect a given matter element in the perturbed configuration to that of its position in the unperturbed background.

For an equation of state of the form $P = P\bigl(\rho, \, T \bigr)$, the Lagrangian pressure perturbation $\Delta P$ arising from a Lagrangian temperature perturbation $\Delta T$ is

\begin{equation} \label{eq: Lagrangian pressure}
    \Delta P = \frac{\partial P}{\partial \rho} \biggr|_T \, \Delta \rho \, + \,  \frac{\partial P}{\partial T} \biggr|_{\rho} \, \Delta T \, .
\end{equation}

\noindent
The large-scale temperature asymmetry described in the previous section is then precisely the means through which one is able to introduce an effective angular dependence in the equation of state - see equations \eqref{eq:deltaTSH Eulerrian} - \eqref{eq:deltaTSH Lagrangian}. What follows is a brief description of the two possible mechanisms previously mentioned which couple the temperature perturbations into an elastic readjustment of the crust: \\

\begin{enumerate}
    
    \item[(I)] \noindent the \citet{Ushomirsky_2000} model, exploiting temperature-dependent electron capture reactions. \\
    
    \item[(II)] \noindent the \citet{Jones_2025} model, exploiting the temperature dependence of the crust's ionic lattice.
\end{enumerate}

\noindent
We now describe each of these mechanisms in turn. 

\subsubsection{Electron capture layer shifts}

The original concept behind the thermal mountain proposed by \citet{Bildsten_1998} suggested that lateral temperature gradients could give rise to so-called ‘wavy' electron capture layers in the crust of \textit{accreting} neutron stars. Accretion of matter onto the star's surface changes the composition of the original crust. The composition, which is the run of the mass number $A$ and atomic number $Z$ of the nuclei, varies with depth. Thermonuclear burning of accreted matter (hydrogen, helium, etc.) via the rp process generates a layer of nuclear ashes on the surface ($A \sim 60 - 100$), which subsequently sink into the crust under the compressive weight of freshly accreted matter. Over time, the original crust is reprocessed through a series of electron captures and pycnonuclear reactions \citep{Sato_1979, Schatz_1999, Haensel_1990a}.

Crucially, the transition from one nuclear species to another, while predominantly density-dependent, also has a small temperature dependence \citep{Bildsten_1998_Cumming}. Exactly how the temperature perturbations translate into a perturbation in the star’s density profile were assumed to enter through a temperature dependence on the electron mean molecular weight $\mu_{\text{e}}$, in a two-parameter EOS whereby $P = P [\rho, \, \mu_{\text{e}}(\rho, T)]$, with $\mu_{\text{e}}$ being equivalent to the threshold energy $E_{\text{cap}}$ for a given electron capture event. Due to their complexity, the interested reader is directed to Appendix A of \citet{Ushomirsky_2000} in order to see the formulae that describe $\Delta P$ due to electron capture. 

From their detailed numerical calculation, when the threshold energy exceeds $90$ MeV, UCB found that the resultant mass quadrupole deep in the inner crust is in excess of $10^{37}$ g cm$^{2}$ (or equivalently $\varepsilon \sim 10^{-8}$). However, capture layers with such large $E_{\text{cap}}$ are absent in modern detailed calculations of the crustal EOS (see Section 2 of \citealt{Jones_2025} for a summary description of modern accreted EOS calculations). The most recent calculation by \citet{Gusakov_2020}, for example, predicts the maximum threshold energy to be $\sim 25$ MeV, and there to be \textit{no} capture layers in the inner crust (i.e. past the neutron drip point) at all. 

In the outer crust, the mass quadrupole itself, and by extension the ellipticity, scales steeply with the threshold energy and can be estimated as \citep{Ushomirsky_2000}

\begin{equation}\label{eq: UCB Fiducial epsilon}
  \Tilde{\varepsilon} = 1.6 \times 10^{-10} \, \bigg( \frac{\delta T /T }{1\%} \biggr) \, \bigg( \frac{E_{\text{cap}}}{30 \, \text{MeV}} \biggr)^3 \, ,
\end{equation}

\noindent 
for a canonical $10$\,km neutron star. Making a simple comparison\footnote{A 90 MeV capture layer would reside in the inner crust, where Equation \eqref{eq: UCB Fiducial epsilon} is not strictly valid. A more detailed comparison can be found in Section 2 of \citet{Jones_2025}.}, this implies that there is a factor $\sim 50$ difference in $\Tilde{\varepsilon}$ between a capture layer with $E_{\text{cap}} = 25$ MeV and one with $E_{\text{cap}} = 90$ MeV. The absence of such high-threshold capture layers in modern EOS calculations clearly reduces significantly the size of the mountains that can be generated via UCB's shifting capture layer model.

\subsubsection{Thermal lattice pressure}
\label{SubSubSec: Crustal Lattice Pressure}

It is expected that relativistic electrons supply the majority of the pressure in the outer crust, while neutrons supply most of the pressure in the inner crust.  To a good approximation, the total pressure is determined by the zero-temperature EOS. However, as discussed in \citet{Jones_2025}, there will be some small thermal correction, increasing the pressure with respect to the zero-temperature approximation. Contributions from the electrons and neutrons are typically modelled in terms of a pressure arising from Fermi \textit{gases} (e.g. \citealt{Chamel_2008}), which presents an immediate problem: in the context of building mountains, a sustained deformation of the crust is required. It is unfortunately likely that any thermally-induced perturbations in these gases would be convected away far too quickly to be relevant for mountain formation.

There is, however, another component of the crustal pressure that is necessarily firmly tied to the elastic phase. More specifically, there is a pressure that is generated directly by the ionic lattice itself, via interaction of the ions with other ions, as well as with the background sea of electrons. This \textit{crustal lattice pressure} has a temperature-dependent piece, and is thus sensitive to any existing anisotropies in the crust. A comparison between the magnitude of the lattice-pressure contribution and the thermal corrections to the electron and neutron pressures is given in \citet{Jones_2025}, who found that these contributions are of comparable magnitude for typical LMXB crust temperatures ($T \sim 10^8 \, \mathrm{K}$).  Note, however, the key distinction that, for the lattice contribution, the ions are anchored to lattice sites: any thermally-induced perturbations cannot be as easily erased by fluid motion, and thus will likely create longer-lived deformations of the crust.  See also the related discussion of a multifluid treatment of the problem, at the end of Section 8.

It is at this point we also note that while in this paper we are considering mountain formation in LMXB systems, another advantage of our crustal lattice pressure mechanism over the UCB model is that it is also relevant in \textit{non-accreting} systems as well. In the present case we take advantage of so-called deep crustal heating (Sec. \ref{Sec: Thermal Structure of Accreting Neutron Stars}) - a mechanism specific to accreting NSs - to warm the crust. Note, however, that in a magnetar, significant amounts of Joule heating due to the dissipation of strong magnetic fields ($B \sim 10^{14}$ G) could also provide the necessary warming effect on the lattice. This, coupled with inevitably strong anisotropic heat conduction from the very same magnetic fields would then drive the required pressure imbalance in much the same way we consider here. Of course, there will also be direct distortions (via Lorentz stresses)  of the star giving conventional magnetic mountains, which for magnetars can be large (see Equation \eqref{eq: Magnetic mountain scaling}).  A comparison of our thermo-elastic mountains and conventional magnetic mountains is given in Section \ref{SubSubSec: Thermal mountains vs. magnetic mountains}.

In any case, while more detail can be found in \citet{Jones_2025}, below we detail the relevant parts of the derivation of the thermal lattice pressure as needed for the calculation of the mass quadrupole in Section \ref{Sec: Quadrupolar Deformations of Accreting Neutron Stars} for LMXB systems.

Utilizing detailed descriptions of the thermodynamic properties of neutron star crusts in \citet{Haensel:2007yy}, the thermal pressure from the lattice can be written as \citep{Jones_2025} 

\begin{equation}\label{eq: Lattice Pressure}
    P_{\text{th}} = \frac{1}{2}\hbar \omega_{\rm{pi}} \biggl[ \frac{ \, \rho (1 - X_{\text{n}})}{ m_{\text{b}}  A} \biggr] \frac{d f_{\text{th}}(\theta)}{d\theta}  \, , 
\end{equation}

\noindent
where $X_{n}$ is the fraction of free neutrons in the inner crust (obtained from the equation of state), $\omega_{\rm{pi}}$ is the plasma ion frequency, $f_{\text{th}}$ is the reduced thermal free energy, and $\theta$ is a `quantum parameter', given by \citep{Baiko_2001}

\begin{equation}\label{eq: Theta}
    \theta \equiv \frac{\hbar}{k_B T} \biggl[ 4 \pi e^2 n_{\text{N}} \frac{Z^2}{m_i} \biggr]^{1/2} \, ,
\end{equation}

\noindent
where $n_{\text{N}}$ is the ion number density, and $m_i$ is the ion mass $m_i = A m_{\text{b}}$. \\

For our calculation, we use the convenient fitting formula for the reduced thermal free energy of a harmonic Coulomb crystal provided by \citet{Baiko_2001}, given by

\begin{equation}\label{eq: Reduced Free Energy}
    f_{\text{th}} = \sum^3_{n = 1} \text{ln}(1 - e^{-\alpha_n \theta}) - \frac{A(\theta)}{B(\theta)} \, , 
\end{equation}

\noindent
with the quantities $A(\theta)$ and $B(\theta)$ given by

\begin{equation}\label{eq: Free energy A}
    A(\theta) = \sum^8_{n = 0} a_n\theta^n .
\end{equation}

\begin{equation}\label{eq: Free energy B}
     B(\theta) = \sum^7_{n = 0} b_n\theta^n + \alpha_6 a_6 \theta^9 + \alpha_8 a_8 \theta^{11} \, ,
\end{equation}

\noindent
where $\alpha_n$, $a_n$ and $b_n$ are sets of constants with values given in Table \ref{tab:Parameters of Free Energy}, reproduced (for convenience) from Table II of \citet{Baiko_2001} for the specific case of a body-centered cubic Coulomb lattice.

\begin{table}
	\centering
	\caption[Parameters for the reduced thermal free energy of a Coulomb crystal]{Parameters needed to calculate the reduced thermal free energy of a body-centred cubic Coulomb lattice. This table is a reproduction of Table II of \citet{Baiko_2001}.}
	\label{tab:Parameters of Free Energy}
    \begin{tabular}{l|cccc}
        \hline
        \multicolumn{1}{c|}{n} & $\alpha_n$    & $a_n$         & $b_n$    \\ \hline
        0                      & -        & 1         & 261.66   \\
        1                      & 0.932446 &  0.1839   & 0.0   \\
        2                      & 0.334547 & 0.593586    & 7.07997   \\
        3                      & 0.265764 & 5.4814 $\times 10^{-3}$  &  0.0    \\
        4                      & -    & 5.01813 $\times 10^{-4}$    & 0.0409484    \\
        5                      & -    & 0.0     & 3.97355 $\times 10^{-4}$   \\
        6                      & 4.757014 $\times 10^{-3}$    & 3.9247 $\times 10^{-7}$     & 5.11148 $\times 10^{-5}$   \\
        7                      & -    & 0.0     & 2.19749 $\times 10^{-6}$   \\
        8                      & 4.7770935 $\times 10^{-3}$     & 5.8356 $\times 10^{-11}$   & -        \\ \hline

    \end{tabular}
\end{table}

Ultimately, we require a form for the thermal lattice pressure that is compatible with the framework of computing the elastic readjustment of the crust described by UCB, and in particular Equation \eqref{eq: Lagrangian pressure}. It is therefore necessary to consider the (Lagrangian) perturbation in the thermal pressure, rather than $P_{\rm th}$ itself. The fractional change in the lattice pressure caused by a fractional temperature (Lagrangian) perturbation $\Delta T / T$ is

\begin{equation} \label{eq: pressure perturbation}
    \frac{\Delta P_{\text{th}}}{P} = \frac{T}{P} \frac{\partial P_{\text{th}}}{\partial T}\biggl |_{\rho} \biggl( \frac{\Delta T}{T} \biggr) \, ,
\end{equation}

\noindent
where the derivative of the thermal pressure is 

\begin{equation}\label{eq: Lattice Pressure Derivative}
    \frac{\partial P_{\text{th}}}{\partial T}\biggl |_{\rho} = \, - \, \frac{1}{2}\hbar \omega_{\rm{pi}} \frac{\theta}{T} \biggl[ \frac{ \, \rho (1 - X_{\text{n}})}{ m_{\text{b}}  A} \biggr] \biggl[ \frac{d^2 f_{\rm{th}}(\theta)}{d\theta^2} \biggr] \, ,
\end{equation}

\noindent
and the second derivative of the reduced thermal free energy $f_{\text{th}}''$ given by \citep{Baiko_2001}

\begin{equation} \label{eq: Second Free Energy Derivative}
\begin{split}
    \frac{d^2f_{\rm{th}}}{d\theta^2} =  -\mathlarger{\sum}_{n=1}^3 & \frac{\alpha^2_n e^{\alpha_n \theta}}{(e^{\alpha_n \theta} - 1)^{2}} - \frac{2A(\theta)(B'(\theta))^2}{B^3(\theta)}  \\ & + \frac{2A'(\theta)B'(\theta)+A(\theta)B''(\theta)}{B^2(\theta)} - \frac{A''(\theta)}{B(\theta)}  \, ,
\end{split}
\end{equation}

\noindent
where the prime in this instance denotes differentiation with respect to the parameter $\theta$.


\section{Models of Accreting Neutron Stars}
\label{Sec: Models of Accreting Neutron Stars}

The composition and EOS of accreting neutron stars was recently calculated by \citet{Fantina_2018, Fantina_2022}, fitted to the energy-density functionals BSk19, BSk20, and BSk21 (for which the composition and EOS of \textit{non-accreted} NSs has already been calculated; \citealt{Goriely_2010, Pearson_2011, Pearson_2012}). Their description improved upon the original works of \citet{Haensel_1990a, Haensel_1990b} (later refined in \citealt{2003HZ, 2008HZ}) in two important regards: (i) the inclusion of nuclear shell effects that predicts a crust with fewer capture layers, and (ii) the outer crust, inner crust, and core of the NS are all treated in a fully thermodynamically consistent way to produce a \textit{unified} EOS. 

Indeed, the original EOS provided by \citet{Haensel_1990b} includes only the parts of the accreted crust where non-equilibrium reactions (i.e. electron capture events) occur. Therefore, to model an accreting NS as a whole, one need `bolt on' the accreted EOS to some other EOS that describes the innermost part the crust, as well as the core. This process, however, can lead to uncertainties in global properties of the star (i.e. mass and radius) of $\sim 5\%$ \citep{Fortin_2016,Suleiman_2021}, and would also lead to significant convergence issues in our calculation of the elastic readjustment of the star (Sec. \ref{SubSec: Analytical Representations of the Accreted Equation of State}). 

In what follows, we shall use the unified accreting EOS of \citet{Fantina_2018, Fantina_2022} to calculate the elastic response of the accreted crust, since the calculation relies heavily on the macroscopic and microscopic properties of both the core and crust (while not suffering from any of the aforementioned thermodynamic inconsistencies). We also note that since we consider the crustal lattice itself for generating pressure and density perturbations, we are effectively insensitive to the number of/locations of the capture layers, which is by far the most prominent difference between the accreted EOS models.

\subsection{Analytical representation of the accreted crustal equation of state}
\label{SubSec: Analytical Representations of the Accreted Equation of State}

A defining feature of an accreted neutron star crust is its layered structure. Throughout the star, it is required that the pressure vary continuously. At a transition $(A, \, Z) \rightarrow (A', \, Z')$, it follows that the pressure in this region must also be continuous \citep{Haensel_1990a}. As a result, there is a density discontinuity, or `jump', every time there is a change of nuclear species at the interface between layers.

These density and compositional discontinuities can introduce an array of difficulties in numerical calculations at points in the crust where they appear\footnote{The location of each interface and associated density jump $\Delta \rho / \rho$ for each of BSk19, BSk20, and BSk21 can be found in the second and fifth columns respectively of Tables A3 - A1 in \citet{Fantina_2018}.}. 

Our method to compute the mass quadrupole requires multiple integrations over the density profile of the crust (Section \ref{Sec: Quadrupolar Deformations of Accreting Neutron Stars}), involving terms (such as the adiabatic index; Section \ref{SubSubSec: Adiabatic Index}) which contain derivatives of $\rho (r)$. Initially, we found that integrating over the discontinuities lead to significant convergence issues, with disparities in the final values for the mass quadrupole obtained via two different methods (see Section \ref{subsec: method of solution}) exceeding an order of magnitude.

We found that an approximation of the EOS may be implemented to circumvent these difficulties. To mitigate issues of convergence, we follow a method outlined by \citet{Potekhin_2013}, who derived a set of 
fully analytical functions approximating the 
BSk19, BSk20, and BSk21 EOSs for \textit{non-accreted} neutron stars.

By introducing two variables $\zeta = \text{log}_{10}(P / \text{dyne \, cm}^{-2})$ and $\xi = \text{log}_{10}(\rho / \text{g \, cm}^{-3})$ - which we shall henceforth label as $\chi$ in order to avoid any possible confusion with the crustal displacement vector $\xi^i$ - the authors were able to parameterise the pressure-density relation as

\begin{equation} \label{eq: Analytic EOS Fit}
\begin{split}
\zeta = \, & \frac{a_1 + a_2 \chi + a_3 \chi^3}{1 + a_4\chi} \{ \text{exp}[a_5(\chi - a_6)] + 1 \}^{-1} \\
& + (a_7 + a_8\chi) \{ \text{exp}[a_9(a_6 - \chi)] + 1 \}^{-1} \\
& + (a_{10} + a_{11}\chi) \{ \text{exp}[a_{12}(a_{13} - \chi)] + 1 \}^{-1} \\
& + (a_{14} + a_{15}\chi) \{ \text{exp}[a_{16}(a_{17} - \chi)] + 1 \}^{-1} \\
& + \frac{a_{18}}{1 + [a_{19}(\chi - a_{20})]^2} + \frac{a_{21}}{1 + [a_{22}(\chi - a_{23})]^2} \, ,
\end{split}
\end{equation}
\\
\noindent
where the values of the parameters $a_i$ for the non-accreting EOSs are given in Table 2 of \citet{Potekhin_2013}. 

In order to parameterise the \textit{accreted} equations of state, we used the Python routine \texttt{curve\_fit} from the scientific computation library \texttt{SciPy} to create a non-linear least squared fit to the analytic function $\zeta$ using tabulated data specific to BSk19, BSk20, and BSk21 for the accreted crust. Initial guesses for $a_{1-23}$ were constructed by fixing them to that of values given in Table 2 of \citet{Potekhin_2013} for the corresponding non-accreted EOSs \citep{Fantina_2022}. The new (optimal) values of $a_i$ were then found by enforcing that the sum of the squared residuals of $\zeta[\chi(\rho), \, a_i] - P$, be minimized. The optimal parameters for the accreted EOSs are given in Table \ref{tab:Analytic Fit Parameters}. 

\begin{table}
\centering
\caption{Values of $a_i$ that parameterise the analytical fit of Eq. (\ref{eq: Analytic EOS Fit}) for the \textit{accreted} equations of state BSk19, BSk20 and BSk21.}
\label{tab:Analytic Fit Parameters}
\begin{tabular}{lccc}
\hline
\hline
$i$ & BSk19                & BSk20       & BSk21                \\ \hline
    & \multicolumn{1}{l}{} & $a_i$       & \multicolumn{1}{l}{} \\
1   & 3.790              & 3.916     & 4.843             \\
2   & 7.461              & 7.436     & 6.989              \\
3   & 0.00759           & 0.00766  & 0.00712           \\
4   & 0.20818            & 0.20799   & 0.19326             \\
5   & 3.913              & 3.588     & 4.078              \\
6   & 12.260               & 12.263     & 12.242             \\
7   & 13.284               & 13.752     & 10.523              \\
8   & 1.3734              & 1.3336    & 1.5900               \\
9   & 3.898              & 3.578     & 4.108             \\
10  & - 13.026            & - 23.342   & - 28.724            \\
11  & 0.9307              & 1.6281   & 2.0854              \\
12  & 5.91              & 4.99     & 4.85              \\
13  & 14.387              & 14.191     & 14.303             \\
14  & 16.652              & 23.575     & 22.880              \\
15  & - 1.0530            & - 1.5222     & - 1.7717            \\
16  & 2.489              & 2.135     & 0.999             \\
17  & 15.405              & 14.980     & 15.329              \\
18  & - 0.026          & - 0.018  & 0.035          \\
19  & 2.25              & 6.67     & 4.64             \\
20  & 11.44              & 11.64     & 11.74              \\
21  & - 0.028          & - 0.031 & - 0.082          \\
22  & 20.3              & 15.0     & 10.0             \\
23  & 14.20              & 14.19     & 14.15              \\ \hline
\hline
\end{tabular}%
\end{table}

\begin{figure*}
\centering
	\includegraphics[width=0.8\textwidth]{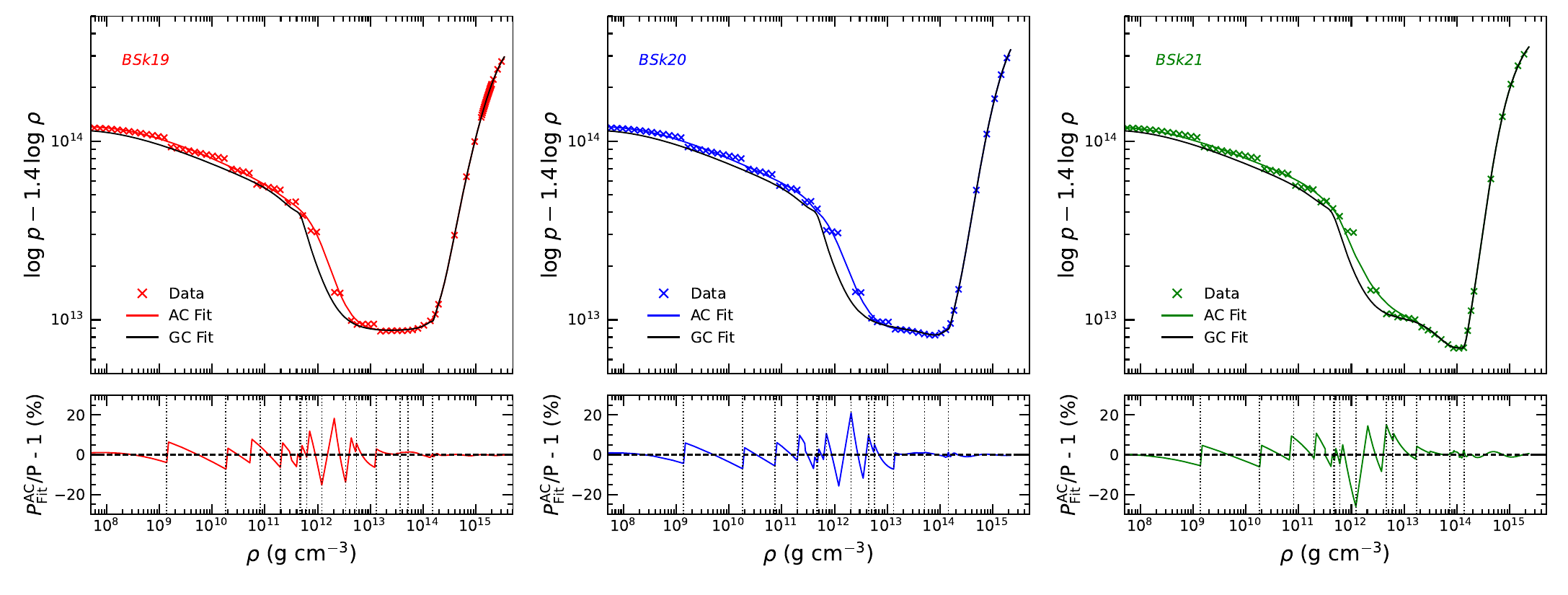}
    \caption{\textit{Upper panels}: Analytical fits to the pressure-density relations predicted by the BSk19 (\textit{left}), BSk20 (\textit{centre}) and BSk21 (\textit{right}) equations of state for `AC' accreted (coloured lines; \citealt{Fantina_2018, Fantina_2022}) and `GC' ground-state non-accreted (black lines; \citealt{Goriely_2010, Pearson_2011, Pearson_2012}) neutron stars. Crosses denote (rarefied) tabular data points for the accreted crust. Coloured lines are obtained via Equation \eqref{eq: Analytic EOS Fit} together with Table \ref{tab:Analytic Fit Parameters}, while black lines are obtained via Equation \eqref{eq: Analytic EOS Fit} and Table 2 in \citet{Potekhin_2013}. \textit{Lower panel}: Relative percentage difference between the tabulated data and analytic fit of the accreted crust. Vertical dotted lines indicate the location of each capture layer from Tables A1 - A3 of \citet{Fantina_2018}.}
    \label{fig:Accreted EOS Analytic Fits}
\end{figure*}

In the upper panels of Figure \ref{fig:Accreted EOS Analytic Fits} we show the BSk19 (left), BSk20 (center), and BSk21 (right) EOSs, together with their analytical representations. For clarity, crosses in the upper panels show (rarefied) tabulated data for the accreted crust (AC), and the colored lines show the analytical fits computed via Equation (\ref{eq: Analytic EOS Fit}) and Table \ref{tab:Analytic Fit Parameters}. To aid comparison, we also include the analytic fits for the non-accreted (ground-state; GC) crust computed from \ref{eq: Analytic EOS Fit}) and Table 2 from \citet{Potekhin_2013}. Rather than just pressure, we plot the function $\zeta - 1.4 \chi$ as a function of the density, since this allows for a better inspection of the fit in the crust, which we care most about. 

Though the tabular data is rarefied, one may easily identify the density jumps with the aid of the vertical dotted lines in the lower panel of Figure \ref{fig:Accreted EOS Analytic Fits}, which indicate the location of each capture layer in density space. The lower panel shows the relative percentage difference between the fits and the tabulated data (of the accreted crust). The average error in the fits is $\approx 0.3 \%$, a similar level of accuracy to the fit of the non-accreted EOSs obtained by \citet{Potekhin_2013}. 

The maximum error we obtain, however, is 26\% (for BSk21), occurring at $\rho = 1.2 \times 10^{12}$ g cm$^{-3}$, corresponding to a density jump $\Delta \rho / \rho = 0.68$ at the location of a pycnonuclear reaction in the inner crust (Table A.3 in \citealt{Fantina_2018}). Such a large discontinuity is challenging for the least-squares fit, which attempts to smoothly interpolate between the two points either side of the jump.  

In what follows, we shall proceed with the fits obtained via Equation \eqref{eq: Analytic EOS Fit}, using the values of $a_i$ given in Table \ref{tab:Analytic Fit Parameters}. We find that the fits do not make any appreciable difference to the hydrostatic and thermal profiles of the star as compared to using only tabulated data (the error is only large in a very narrow range either side of the discontinuities at low density, and the average error in the fits is $\lesssim 1\%$). Any slight losses in accuracy in the thermal calculation, are, in any case, far outweighed by the fact that we achieve much better agreement in the value of $Q_{22}$ (the quantity we are most interested in) obtained from Equations (\ref{eq: Mass Quadrupole 1}) - (\ref{eq: Mass Quadrupole 2}) when using the analytical fit as compared to tabulated data. We will return to this issue later in Section \ref{subsec: method of solution}.

\subsection{Microphysical structure}
\label{SubSec: Microphysical Structure}

\subsubsection{Adiabatic index}
\label{SubSubSec: Adiabatic Index}

From the analytical representations \eqref{eq: Analytic EOS Fit}, differentiation with respect to the mass density $\rho$ allows us to compute the adiabatic index $\Gamma$. For an equation of state of the form $P = P(\rho, \, T)$, this should be done at fixed temperature. In our case however, we simplify the problem by noting that the BSk EOSs are computed at \textit{zero temperature} \citep{Fantina_2018, Fantina_2022}. Given that we seek to only introduce temperature dependence on the equation of state as a perturbation on the zero-temperature EOSs to begin with, we may assume that

\begin{equation}\label{eq: Adiabatic Index}
    \Gamma =  \frac{\partial \, \text{ln}  \,P}{\partial \, \text{ln}  \,\rho} \biggl|_{T} \approx \frac{\text{d ln} \, P}{\text{d ln}  \,\rho} \, . 
\end{equation}

\noindent
It is this form of the adiabatic index that we will include when forming the perturbation equations of the crust in Section \ref{Sec: Quadrupolar Deformations of Accreting Neutron Stars}.

\subsubsection{Shear modulus}
\label{SubSubSec: Shear Modulus}

The crust of a neutron star is usually assumed to be an isotropic, body-centered cubic Coulomb crystal, with an effective shear modulus $\mu$. To further complete the picture of our accreted crust, we use the result of \citet{Zdunik_2008}, obtained from a previous calculation by \citet{Ogata_1990}, which we write as

\begin{equation}\label{eq: Shear Modulus}
\begin{split}
    \mu = 7.8 \times 10^{28} \, \biggl( & \frac{\rho}{10^{13} \, \text{g cm}^{-3}} \biggr)^{4/3} \, \cdot \\
    & \hspace{0.7cm} \biggl( \frac{10^3 (1 - X_n)}{A} \biggr)^{4/3} \, \biggl( \frac{Z}{40} \biggr)^2 \, \text{dyne cm}^{-2} \, .
\end{split}
\end{equation}

This formula is used to describe $\mu$ everywhere in the accreted crust. This choice is only an approximation, however, since Equation \eqref{eq: Shear Modulus} is (strictly speaking) only valid for point-like nuclei \citep{Zdunik_2008}. The deepest layers of the crust also likely contain a series of finite-sized `nuclear pastas'  which should, in principle, be modelled differently (see e.g. \citealt{Caplan_2018} and references therein). Since the BSk EOSs assume purely spherical nuclei exist down to the crust-core transition \citep{Pearson_2012,Fantina_2022} we too use Equation \eqref{eq: Shear Modulus} for the deepest layers of the crust, and do not account for the existence of pasta phases. 

The shear modulus is typically small relative to the pressure ($\mu / P \sim 10^{-3} - 10^{-2}$), which is indicative of how large a deformation can be supported by the crust \citep{Ushomirsky_2000}. A larger shear modulus implies the crust is more rigid, and thus resistive to stresses generated within the crustal lattice. The strain exerted on the crust as a result of the quadrupolar magnetic field, and the associated deformation, will be calculated in Section \ref{Sec: Quadrupolar Deformations of Accreting Neutron Stars}.

\section{Hydrostatic structure}
\label{Sec: Hydrostatic Structure}

For a particular equation of state, the hydrostatic structure of neutron stars is described by the well-known Tolman-Oppenheimer-Volkoff (TOV) equation

\begin{equation}\label{eq: TOV equation}
    \frac{dP}{dr} = \, - \, \frac{Gm}{r^2}\rho \, \biggl(1 + \frac{p}{\rho c^2} \biggr) \biggl(1 + \frac{4\pi r^3 p}{m c^2} \biggr) \biggl(1 - \frac{2Gm}{r c^2} \biggr)^{-1} \, , 
\end{equation}

\noindent
together with

\begin{equation}\label{eq: mass derivative}
    \frac{dm}{dr} = 4\pi r^2 \rho \, . 
\end{equation}

\noindent
These structure equations, appended with the analytical EOSs described in Section \ref{SubSec: Analytical Representations of the Accreted Equation of State}, are solved by integrating outward from some chosen central density $\rho_c$, until the pressure drops to zero. In our analysis, we consider three NS models which have the properties listed in Table \ref{tab:Model properties}, obtained by integrating Equations (\ref{eq: TOV equation}) - \eqref{eq: mass derivative} using the values of $\rho_c$ listed in column two. An example schematic cross-sectional view of the accreted crust obtained from the TOV solution for the $1.4$ $M_{\odot}$ neutron star assuming the BSk20 EOS can be seen in Figure 2 of \citet{Hutchins_2023}. 

\begin{table*}
\centering
\caption{Central density $\rho_c$, mass $M$, crustal mass $M_{\text{crust}}$, radius $R$, crust thickness $R_{\text{crust}}$, and crust-core transition densities for three neutron stars assuming each of the equations of state BSk19, BSk20, and BSk21.}
\label{tab:Model properties}
\begin{tabular}{l|ccccccc}
\hline
\hline
EOS & $\rho_c$ {[}10$^{15}$ g cm$^{-3}${]} & $M$ {[}$M_{\odot}${]} & $M_{\text{crust}}$ {[}$M_{\odot}${]} & $R$ {[}km{]} & $R_{\text{crust}}$ {[}km{]} &  $\rho_{\text{crust-core}}$ {[}10$^{14}$ g cm$^{-3}${]} \\ \hline
BSk19 & 1.321 & 1.40 & 0.014 & 10.80 & 0.83 & 1.497 \\
BSk20 & 0.924 & 1.40 & 0.018 & 11.81 & 1.00 & 1.444 \\
BSk21 & 0.732 & 1.40 & 0.018 & 12.64 & 1.11 & 1.367 \\ \hline \hline
\end{tabular}
\end{table*}

For simplicity, we wish to avoid solving the equations which determine the elastic readjustment of the crust in full general relativity. Instead, we would like to remain in a Newtonian setting so as to focus on implementing a realistic source term that has, until now, been absent in the literature. Whilst in future work we would seek to extend the formalism presented here into a fully general relativistic framework; for now, we simply acknowledge the results of \citet{Gittins_2021_relativity}. They found that the \textit{maximum} mountain the crust could sustain due to an \textit{unmodelled} thermal pressure perturbation $\delta P_{\text{th}}$ changes by a factor $\sim 2$ when generalising the scheme to compute mountain sizes from a Newtonian framework \citep{Gittins_2021} into general relativity.


When computing the mass quadrupole moment, we shall therefore map our solution from the relativistic TOV equation to a Newtonian star by reinterpreting the TOV radial coordinate, pressure and energy density as simply their Newtonian counterparts. In this sense, the Newtonian mass density is just the relativistic energy density divided by $c^2$. 

Such a choice has two fundamental consequences. On the one hand, we preserve the original definition of the equation of state (namely the pressure-density relation), but on the other inevitably produce a star whose local acceleration due to gravity $g$ does not quite satisfy Poisson’s equation for gravity, and thus does not match the value that Newtonian theory would prescribe.

As a result, we choose to proceed in the following way. In Newtonian gravity, the equation of hydrostatic balance may be rearranged to give 
\begin{equation}\label{eq: Newtonian Hydro}
    g(r) = \, - \, \frac{1}{\rho} \frac{\text{d}P}{\text{d}r}.
\end{equation}

\noindent
Rather than call $g$ the `gravitational acceleration', defined by $g(r) = Gm(r) / r^2$ explicitly, we shall instead define it in terms of Equation \eqref{eq: Newtonian Hydro}, where the quantity $dP/dr$ in the RHS of Equation \eqref{eq: Newtonian Hydro} is given by the RHS of the TOV equation (\ref{eq: TOV equation}), which constrains the hydrostatic structure of the NS as modelled by general relativity. In practice, if one compares Equation \eqref{eq: Newtonian Hydro} with the RHS of Equation (\ref{eq: TOV equation}), this `effective-acceleration' is equivalent to assuming that
\begin{equation} \label{eq: Hybrid G}
    g_{\text{eff}}(r) = \, - \, \frac{G m(r)}{r^2} \biggl(1 + \frac{P(r)}{\rho(r) c^2} \biggr) \biggl(1 + \frac{4\pi r^3 p(r)}{m(r) c^2} \biggr) \biggl(1 - \frac{2Gm(r)}{r c^2} \biggr)^{-1} \, . 
\end{equation} 

The quantity $g_{\text{eff}}$ is not a gravitational acceleration, but rather a relativistic analogue that we have created that will differ from the Newtonian $g$ by a factor related to the compactness of the star.  In fact, we find that this procedure leads to discrepancies in the value of $g$ obtained from Newtonian theory by $\approx 30\%$.  We feel this is an acceptable price to pay, given the great simplifications that it allows for the thermal and elastic problems that are to follow.  Of course, in the Newtonian limit, Equation \eqref{eq: Hybrid G} reduces to just $g(r) = Gm(r) / r^2$, at which point one would recover the Newtonian gravitational acceleration.

This result may then be used to compute the radial derivative of $g_{\text{eff}}$ (required to solve the equations which determine the elastic readjustment of the crust; Appendix \ref{Subsec A1: The perturbation equations}) as

\begin{equation} \label{eq: derivative Hybrid G}
    \frac{dg_{\text{eff}}}{dr} = \, - \, \frac{1}{\rho^2} \biggl( \rho \frac{d^2P}{dr^2} \, - \, \frac{dP}{dr}\frac{d\rho}{dr} \biggr) \, .
\end{equation}

\section{Thermal Structure of Accreting Neutron Stars}
\label{Sec: Thermal Structure of Accreting Neutron Stars}

As described in Section \ref{Sec: Hydrostatic Structure}, we compute the star's thermal profile within a Newtonian (rather than relativistic) setting.  We follow closely the procedure outlined in \citet{Hutchins_2023} (hereafter HJ), making use of the following set of coupled  ODEs

\begin{equation} \label{eq:dLdr}
 \frac{dL}{dr} = 4 \pi r^2 (Q_{\rm{h}} - Q_{\nu}) \, ,
\vspace*{3mm}
\end{equation}

\begin{equation} \label{eq:dTdr}
 \frac{dT}{dr} = -\frac{1}{\kappa} \frac{L}{4 \pi r^2} \, ,
\vspace*{3mm}
\end{equation}

\noindent
that are derived from the equation for energy conservation and Fourier's law respectively, for a spherically symmetric star. In Equations \eqref{eq:dLdr} - \eqref{eq:dTdr}, $L$ is the local luminosity, related to the thermal heat flux $\boldsymbol{F}$ via $L = 4 \pi r^2 \, F$ (with $F = |\boldsymbol{F}|$), $Q_{\rm{h}}$ is the local heat energy deposited in the accreted crust per accreted nucleon per unit volume, $Q_{\nu}$ is the local energy loss due to neutrino emission, and $\kappa$ is the thermal conductivity. 

Solving these ODEs requires an accurate description of various microscopic properties, including relevant heat generation, neutrino cooling, and heat transport mechanisms specific to accreting neutron stars. We direct the interested reader to Section 3 of HJ for a detailed description of these processes (as well as references therein). The different neutrino processes and heat transport mechanisms we include in our model may be found in their Tables 2 and 5.

The heat generated in the crust is generally divided into two categories, the so-called \textit{deep} crustal heating and \textit{shallow} crustal heating mechanisms. Deep crustal heating, concentrated in the crustal layers $\sim 10^9 - 10^{13}$ g cm$^{-3}$ occurs as a result of non-equilibrium reactions (electron captures, pycnonuclear reactions and neutron emissions - listed in the third column of Tables A1 - A3 in \citealt{Fantina_2018}). The total amount of heat released per accreted nucleon in each of the BSk19, BSk20, and BSk21 models is 1.54, 1.62, and 1.65 MeV respectively, with the exact amount of heat released due to each individual reaction given in the final column of Tables A1 - A3 in \citet{Fantina_2018}. 

Modelling of light curves of transiently accreting LMXBs however requires (additional) shallow crustal heating - large amounts of heat in the outer portion of the crust ($\rho < 10^{10}$ g cm$^{-3}$) - in order to explain observed temperatures during periods of quiescence. The sources of this additional heating are currently unknown, but is typically in the range 1 - 2 MeV per accreted nucleon (though it can exceed 10 MeV in some cases; e.g. \citealt{Deibel_2015}). 

The rate of accretion onto the NS itself also varies from source to source (see e.g. Table 2 of \citealt{galloway_goodwin_keek_2017}). Since the heat release is typically measured per accreted nucleon (as above), one should expect that a greater observed accretion rate will result in a hotter crust. 

\begin{figure*}
    \centering
	\includegraphics[width=\textwidth]{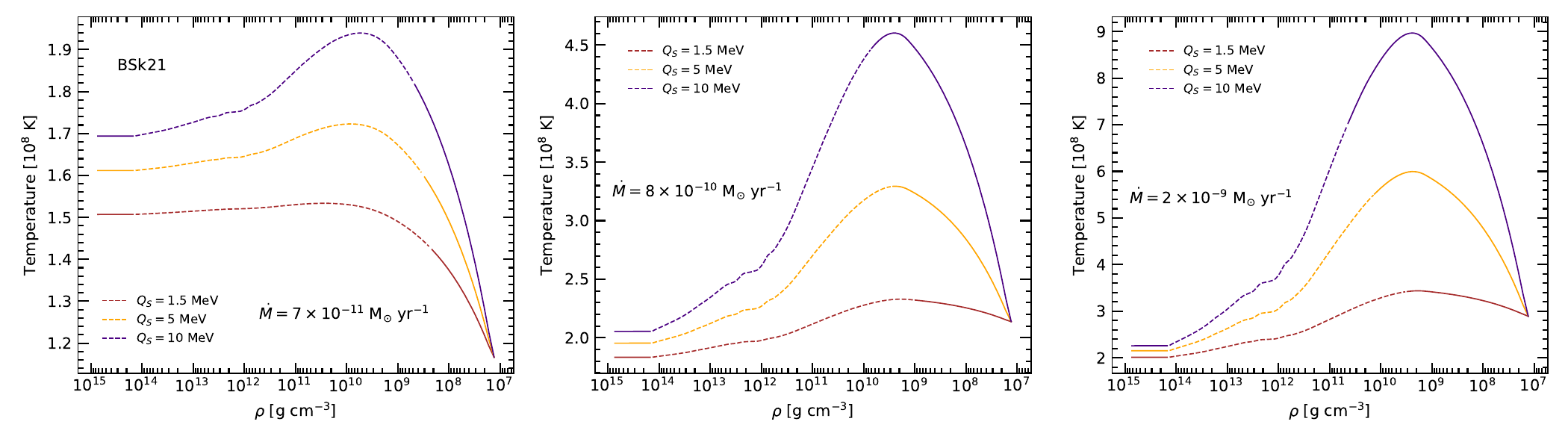}
    \caption{Temperature profiles (in units of 10$^8$ K) of neutron stars assuming the BSk21 equation of state as a function of density, accreting at $\dot{M} = 7 \times 10^{-11}$ M$_{\odot}$ yr$^{-1}$ (\textit{Left}), $\dot{M} = 8 \times 10^{-10}$ M$_{\odot}$ yr$^{-1}$ (\textit{center}), and $\dot{M} = 2 \times 10^{-9}$ M$_{\odot}$ yr$^{-1}$ (\textit{Right}). We also consider different amounts of shallow crustal heating, ranging from 1.5 - 10 MeV, as indicated in the legend. The solid lines indicate fluid regions of the star, while dashed lines indicate regions where the star forms a solid Coulomb lattice. The crust begins at the point where the Coulomb parameter $\Gamma_{\text{Coul}} = 175$ - assuming a one component plasma; Eq. \eqref{eq:Coulomb to Thermal}, and ends at the crust-core transition (Table \ref{tab:Model properties}).}
    \label{fig:BSk20 Background Temperature Profiles}
\end{figure*}

We show in Figure \ref{fig:BSk20 Background Temperature Profiles} the background thermal structure of a number of $1.4$ M$_{\odot}$ neutron stars assuming the BSk21 EOS obtained from solving Equations \eqref{eq:dLdr} - \eqref{eq:dTdr}. Each curve represents a NS with the hydrostatic properties listed in the bottom row of Table \ref{tab:Model properties}, which is steadily accreting in the interval $ \langle \dot{M} \rangle \sim 10^{-11} - \, 10^{-9} \, M_{\odot} \, \text{yr}^{-1}$, with a shallow heating term $Q_{\text{S}}$ that varies in the range 1.5 - 10 MeV, as outlined in the figure. 

\begin{figure*}
    \centering
	\includegraphics[width=\textwidth]{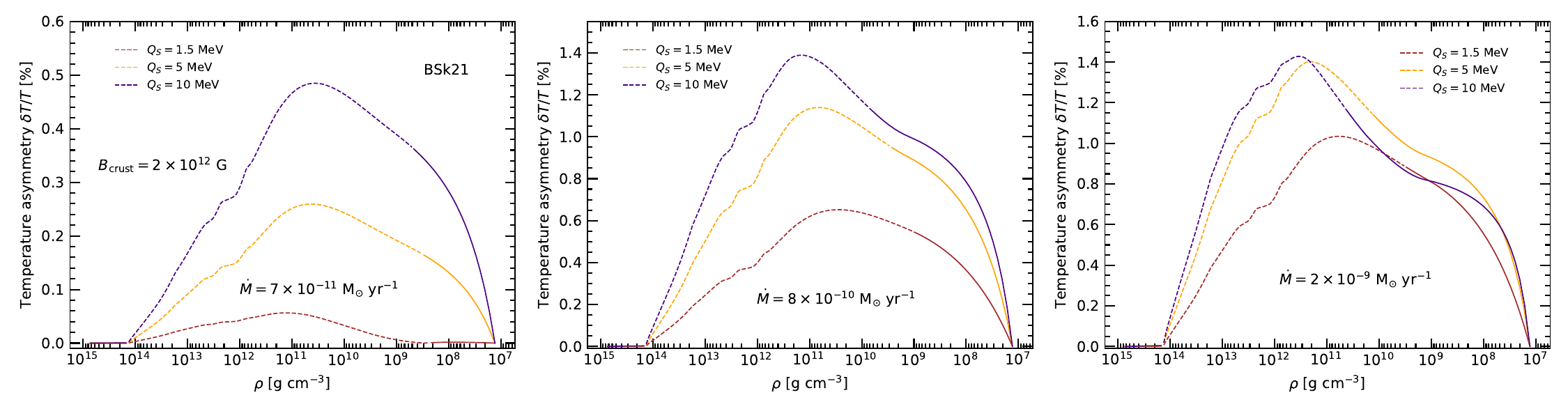}
    \caption{Magnitude of the fractional temperature perturbation $\delta T / T$ inside weakly magnetised neutron stars for the NS configurations considered in Fig. \ref{fig:BSk20 Background Temperature Profiles}. In this figure we assume a $B = 2 \times 10^{12}$ G internal \textit{crustal} toroidal magnetic field.}
    \label{fig:BSk20 Perturbed Temperature Profiles Crust}
\end{figure*}

\begin{figure*}
    \centering
	\includegraphics[width=\textwidth]{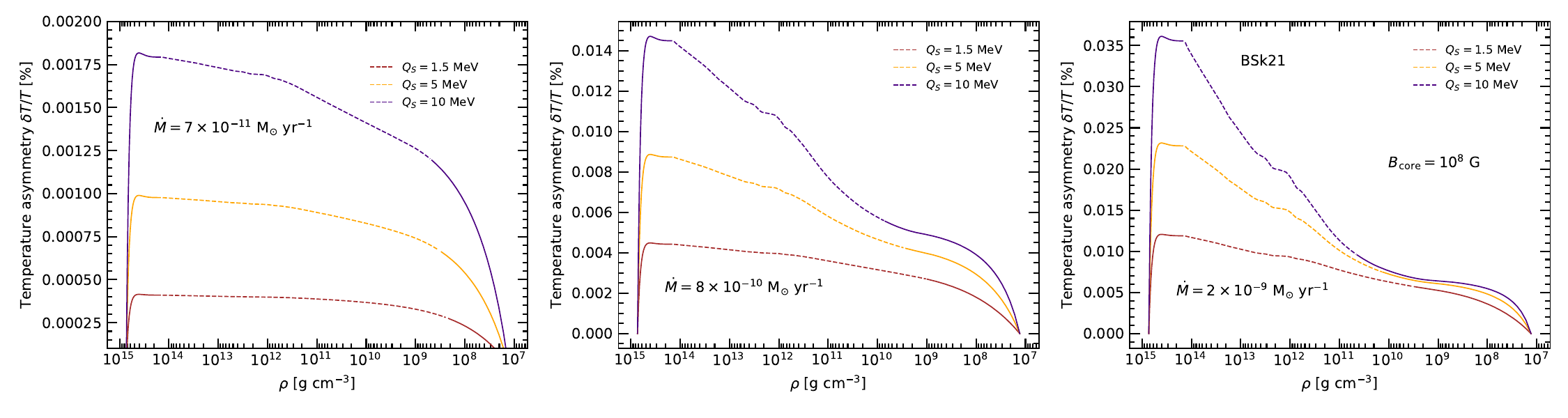}
    \caption{Same as Fig. \ref{fig:BSk20 Perturbed Temperature Profiles Crust} but for a $B = 10^{8}$ G internal \textit{core} toroidal magnetic field.}
    \label{fig:BSk20 Perturbed Temperature Profiles Core}
\end{figure*}

The neutron star is composed of a solid crust bounded by a fluid core and fluid ocean. We have used dashed lines in Figure \ref{fig:BSk20 Background Temperature Profiles} to indicate the solid region of the crust, which can be seen to vary in size (in terms of its diameter) depending on the temperature. It is this region between the crust-core and crust-ocean interfaces that will serve as the computational domain of the elastic calculation (Sec. \ref{subsec: method of solution}), since it is only the solid region of the crust than can support shear stresses ($\mu \equiv 0$ in any fluid regions).

We make the conventional assumption that the solid crust begins at the crust-ocean interface, when the ratio of Coulomb energy to thermal energy,

\begin{equation} \label{eq:Coulomb to Thermal}
\Gamma_{\text{Coul}}  = \frac{Z^2e^2}{k_BT} \biggl( \frac{4\pi n_{\rm{N}}}{3} \biggr)^{1/3} \, ,
\vspace*{3mm}
\end{equation}

\noindent
exceeds the canonical value 175 for a one component plasma \citep{Haensel:2007yy}, where $n_{\rm{N}}$ is the ion number density given by

\begin{equation}
    n_{\rm{N}} = n_{\rm{b}} \frac{1 - X_{\rm{n}}}{A} \, ,
\end{equation}

\noindent
where $n_{\rm{b}}$ is the baryon number density. The solid crust is assumed to end at the crust-core transition, which we give for each of BSk19, BSk20, and BSk21 in the final column of Table \ref{tab:Model properties}.

\section{Perturbations in the Thermal Structure}
\label{Sec: Perturbations in the Thermal Structure}

Thermal mountains are sourced by large-scale temperature gradients within the star. To generate the asymmetry, HJ - following \citet{Osborne_2020} - exploited the likely weak internal magnetic fields of neutron stars in LMXBs to introduce perturbatively some anisotropy in the heat conduction via charged relativistic electrons (responsible for the majority of the heat conduction in NSs) and non-relativistic muons (if available).

The thermal heat flux in the presence of a magnetic field was derived by \citet{Yakovlev_1980}. For a magnetic field $\textbf{B} = \text{B}\textbf{b}$, the heat flux carried by charged species $\text{x}$ is written as

\begin{equation} \label{eq:Perturbed Flux}
    \textbf{F}_{\text{x}}  = - \, \kappa^{\perp}_{\text{x}} \bigl[ \nabla T + (\omega_B^\text{x} \tau_{\text{x}})^2 (\textbf{b} \cdot \nabla T) \textbf{b} + \omega_B^\text{x} \tau_{\text{x}} (\textbf{b} \times \nabla T) \bigr] \, ,
    \vspace*{3mm}
\end{equation}

\noindent
where $\kappa^{\perp}$ is the magnitude of the thermal conductivity perpendicular to the magnetic field, $\textbf{b}$ is the unit vector of the magnetic field (pointing in the direction of the magnetic field) and $\omega_B^\text{x} \tau$ is known as the `magnetisation parameter'. The induced phenomenon is a thermal analogue of the classical Hall effect, with the magnetic field deflecting a piece of the thermal heat flux orthogonal to both the temperature gradients and the local magnetic field, as can be seen from the final term in Equation \eqref{eq:Perturbed Flux}.

The ratio of the conductivities parallel and perpendicular to the magnetic field is given in terms of the magnetisation parameter

\begin{equation}\label{eq:MagParameter}
    \frac{\kappa^{\parallel}_{\text{x}}}{\kappa^{\perp}_{\text{x}}} = 1 + \bigl[\omega^{\text{x}}_B \tau_{\text{x}} (T) \bigr]^2 \, ,
    \vspace*{3mm}
\end{equation}

\noindent
where $\tau_{\text{x}}$ is the collision relaxation time of a heat carrier $\text{x}$ (see Sec. 3.4 of \citealt{Hutchins_2023}), and $\omega_B$ is the gyrofrequency of the carrier, defined as

\begin{equation}\label{eq:Gyrofrequency}
    \omega_B^{\text{x}} = \frac{eB}{m^{\ast}_{\text{x}} c} \, ,
    \vspace*{2mm}
\end{equation}

\noindent
where $m^{\ast}$ is the effective mass of particle $\text{x}$. The thermal conductivity is suppressed in the direction perpendicular to the magnetic field as $B$ increases, leading to a proportionate diminishing of the heat flow orthogonal to the magnetic field and generating the required large-scale asymmetry in the star's global thermal structure. 

For the comparatively weak internal fields of LMXBs, in the limit $\omega^\text{x}_B \tau_\text{x} \ll 1$, HJ linearised Equation \eqref{eq:Perturbed Flux} in both the temperature perturbation $\delta T$ and the magnetisation parameter $\omega_B \tau$. The spherical symmetry of the background was used to describe the dependence of all perturbed quantities on the spherical polar angles $(\theta, \phi)$ using spherical harmonics - e.g. $\delta T = \delta T_{\ell m}(r) Y_{\ell m} (\theta, \phi)$ - such that they were able to derive a set of first-order coupled ODEs to describe the perturbed thermal structure as

\begin{equation} \label{eq:Perturbed Temp ODE}
    \frac{d \delta T_{\ell m}}{dr}  = - \frac{1}{\kappa} \Biggl[ \Biggl( \frac{d \kappa^0_e}{dT} + \frac{d \kappa^0_{\mu}}{dT} + \frac{d \kappa^0_n}{dT} \biggr) \frac{dT}{dr} \delta T_{\ell m}  + U_{\ell m} \Biggr] \, ,
    \vspace*{3mm}
\end{equation}

\begin{equation}\label{eq:Perturbed U ODE}
    \frac{d U_{\ell m}}{dr}  = \frac{d Q_{\nu}}{dT} \delta T_{\ell m}  - \frac{2}{r} U_{\ell m} + \frac{1}{r} l(l+1) V_{\ell m} \, ,
    \vspace*{3mm}
\end{equation}

\noindent
where $\kappa^0_\text{x}$ (with x = n, e, $\mu$) denotes the `non-magnetic' scalar conductivity contributions from neutrons, electrons, and muons respectively (see Table 5 in HJ), and $U_{lm}$ is the radial heat flux perturbation introduced by the magnetic field (Eq. (30) and Eq. (41) in HJ). The (unknown) quantity $V_{\ell m}$ is given by the algebraic expression

\begin{equation} \label{eq:V_lm}
 V_{lm} = \,  \biggl(\kappa^0_{e}\Tilde{\omega} \, \tau_{e} + \kappa^0_{\mu} \Tilde{\omega} \, \tau_{\mu} \biggr) \, \Psi_{lm} \frac{dT}{dr} - \frac{1}{r} \kappa^0 \delta T_{lm} \, ,
\end{equation}

\noindent
where $\Psi_{\ell m} \, (r)$ is a scalar function that describes the form of a \textit{toroidal} magnetic field. 

Magnetic fields within neutron stars are expected to have both poloidal $\textbf{B}_{\text{pol}} = - \nabla \times (\textbf{r} \times \boldsymbol{\nabla} \Phi)$ and toroidal $\textbf{B}_{\text{tor}} = - \textbf{r} \times \boldsymbol{\nabla} \Psi$ components \citep{Radler_2001}. However, as first noted in \citet{Osborne_2020}, the poloidal component $\Phi$ induces a purely toroidal perturbation in the axial component of the heat flux, and therefore produces no perturbations in the temperature profile of the star (see Section 4.1 of HJ for a full description). 

HJ therefore considered the temperature perturbations induced by a toroidal magnetic field only, and in three different scenarios: where the magnetic field (i) permeates the entire star; (ii) permeates only the outer core, due to partial expulsion of the magnetic field in the presence of proton superconductivity; and (iii) is confined to only the crust, as a result of complete expulsion of the magnetic field from the core. The functional form of the scalar function $\Psi_{\ell m} \, (r)$ in each scenario is discussed in Section 4.3 of HJ - see specifically their Equations (48) and (49). 

In Figures \ref{fig:BSk20 Perturbed Temperature Profiles Crust} -  \ref{fig:BSk20 Perturbed Temperature Profiles Core} we show the magnitude of the fractional temperature perturbation $\delta T / T$ (as a percentage) for the background temperature profiles considered in Figure \ref{fig:BSk20 Background Temperature Profiles}. Firstly, in Figure \ref{fig:BSk20 Perturbed Temperature Profiles Crust} we show the fractional temperature asymmetry induced by a $B = 2 \times 10^{12}$ G  \textit{crustal} magnetic field. Such a choice reflects the findings in Section 4.4 of HJ, that 1\% asymmetry may be achieved if the internal magnetic field was assumed to be $\sim 10^{12}$ G. 

In Figure \ref{fig:BSk20 Perturbed Temperature Profiles Core}, on the other hand, we show the fractional temperature asymmetry induced by a $B = 10^8$ G  \textit{core} magnetic field. Due to the perturbative nature of our approach, core magnetic fields are restricted to $B \lesssim 10^8$ in order to satisfy the condition that $\omega_B \tau \ll 1$ is obeyed everywhere so that the perturbation equations \eqref{eq:Perturbed Temp ODE} - \eqref{eq:V_lm} remain valid (this is discussed in Section 4.4 of HJ). We do note, however, that core magnetic fields are potentially (much!) stronger than this, and it is therefore possible that a non-perturbative calculation (which would allow for stronger magnetic fields) could yield stronger asymmetries than currently predicted. We also note that, for \textit{crustal} magnetic fields, HJ found that the condition $\omega_B \tau \ll 1$ was satisfied everywhere for magnetic field strengths $B \lesssim 10^{13}$ G. 

The minimum temperature asymmetry of $\sim 1\%$ required by UCB to generate significant mass quadrupoles from capture layer shifts can be achieved for strongly accreting neutron stars with crustal magnetic fields of the order $10^{12}$ G. Whether or not LMXBs have such strong internal magnetic fields is unknown, though it is worth bearing in mind that inferences of the strength of the external dipolar magnetic field of LMXBs is $\lesssim 10^9$ G.

\section{Quadrupolar Deformations of Accreting Neutron Stars}
\label{Sec: Quadrupolar Deformations of Accreting Neutron Stars}

To compute the mass quadrupole moment of our magneto-thermo-elastic mountains, we make use of the perturbation formalism derived by UCB (their Section 4). Rather than reproduce the entire derivation, we detail the most relevant aspects in Appendix \ref{Sec: Perturbation formalism of the crustal displacements}, so that the reader may see how the various quantities defined in Sections \ref{Sec: Models of Accreting Neutron Stars} - \ref{Sec: Perturbations in the Thermal Structure} (the shear modulus $\mu$, the temperature perturbations $\delta T$, etc.) fit into our mountain formation scenario. We also provide some additional context on the choice of boundary conditions in Appendix \ref{Subsec A2: Boundary Conditions}, which we also take from UCB.

In what follows, there is one subtle (but nonetheless important) distinction to note between our calculation and that of UCB's. The perturbation equations we solve in \eqref{eq: z1} - \eqref{eq: z4} for the specific case of temperature anisotropy are semantically slightly different.  In our calculation, the temperature perturbations described in Section \ref{Sec: Perturbations in the Thermal Structure} (denoted as $\delta T$) were computed for a \textit{fixed crust}, which is to say that we have not yet allowed for any elastic readjustment of the crust. Now, however, we are truly seeking to quantify the response of fluid elements in the crust to these temperature perturbations; solving for the \textit{Lagrangian} displacement field $\xi^i$ that brings the crust back into equilibrium. 

As such we identify these temperature perturbations with the \textit{Lagrangian} temperature perturbations of the elastically deformed star. We do this, as we wish our final solution to be self-consistent with respect to both thermal and elastic perturbations \textit{after the elastic readjustment of the star}. We spell this out, since, although this is precisely the procedure followed by UCB when considering perturbations due to lateral \textit{composition} gradients (i.e. $\Delta \mu_{\text{e}}$), the authors followed a different logic for their temperature perturbations. More specifically, UCB chose to identify fixed-crust temperature perturbations with the \textit{Eulerian} perturbations of the elastically deformed star. Why such a choice was made is unclear to us, and is in our opinion, incorrect. This choice by UCB leads to an additional term in their version of equations \eqref{eq: z1} - \eqref{eq: z4} that we do not include. We demonstrate mathematically how this subtle difference in definition leads to the additional term in Appendix \ref{Subsec A1.1: Boundary Conditions}.

\subsection{Method of solution} \label{subsec: method of solution}

To solve the crustal equations \eqref{eq: z1} - \eqref{eq: z4} we again make use of the \texttt{solve\_BVP} function from the \texttt{SciPy} library, with initial guesses constructed from an order of magnitude estimate of the results of UCB (taken from their Figures 10 - 12). From these solutions, we then require a method to obtain the density perturbation $\delta \rho$ such that we may calculate the mass quadrupole moment as per Equation \eqref{eq: Mass Quadrupole}. Such a method was also derived in UCB, which we reproduce for convenience in Appendix \ref{Subsec A3: Obtaining the quadrupole moment}. 

To test the accuracy of the solutions, we used both equations \eqref{eq: Mass Quadrupole 1} and \eqref{eq: Mass Quadrupole 2} in order to calculate the quadrupole deformation, which we obtain via a simple Simpson's method (taken from the \texttt{SciPy} library) using the solutions to equations \eqref{eq: z1} - \eqref{eq: z4} as computed via \texttt{solve\_BVP}. As we first mentioned in Section \ref{SubSec: Analytical Representations of the Accreted Equation of State} however, while mathematically equivalent, initially we could not get an acceptable level of agreement between the two methods when using the tabulated EOS data. We attribute this to having to integrate multiple times over the density profile of the crust, which, as noted in Section \ref{SubSec: Analytical Representations of the Accreted Equation of State}, contains large density discontinuities at each nuclear transition. When using the analytical fit, we found agreement between the two methods to be in the region $\lesssim 1 \%$ for all of our NS models.

To provide further confidence in our results, we also note that it is possible to recast Equation \eqref{eq: Mass Quadrupole 2} as an additional ODE, taking derivatives on both sides such that 

\begin{equation}\label{eq: Mass quadrupole ODE}
    \begin{split}
     \frac{d  Q_{22}(r)}{d \, \text{ln} \, r} = - \frac{\rho}{\Tilde{V}} \biggl\{& 6z_4 - 2\frac{\mu}{p} \biggl[ 2 \frac{d z_1}{d \, \text{ln} \, r} + 6z_3 \biggr] \\ & + (6 - \Tilde{U})(z_2 - \Tilde{V}z_1) \biggr\} r^5 \, ,
\end{split}
\end{equation}

\noindent
where $Q_{22}(r)$ is the \textit{cumulative} mass quadrupole, with the \textit{total} mass quadrupole being $Q_{22} = Q_{22} (r = r_{\text{crust-ocean}})$, the value of $Q_{22}(r)$ at the crust-ocean interface. 

This additional ODE may be solved simultaneously in conjunction with the perturbation equations (\ref{eq: z1}) - (\ref{eq: z4}) once an additional boundary condition has been specified (since the system becomes a total of five equations rather than just four). A natural choice is the condition that $Q_{22} (r = r_{\text{crust-core}}) = 0$, since at the transition from the base of the solid crust to the liquid core the shear stresses must vanish. 

The benefit of this approach is that it eliminates any potential truncation errors when numerically integrating equations \eqref{eq: Mass Quadrupole 1} - \eqref{eq: Mass Quadrupole 2} with the Simpson method, which is a possible cause for the discrepancies observed between the two methods. Instead, errors are controlled internally by the \texttt{solve\_BVP} function itself on the same mesh used to compute $z_1 - z_4$, where the desired tolerance of the solution can be manually controlled via \texttt{solve\_BVP} algorithm.

Encouragingly, we find agreement between the quadrupole moment obtained from \texttt{solve\_BVP} and the one obtained via the Simpson method to be $\lesssim 0.1\%$. This suggests that a truncation error is not the cause of the remaining differences in $Q_{22}$ obtained via Equation \eqref{eq: Mass Quadrupole 1} and Equation \eqref{eq: Mass Quadrupole 2} with the analytical EOS representations. Instead, we attribute the difference to round-off errors in computing Equation \eqref{eq: Mass Quadrupole 1}, which essentially involves subtracting two large numbers (in excess of $10^{30}$) to get a single, smaller number. This was a similar problem encountered by UCB themselves, where it was noted that `round-off errors [on Equation \eqref{eq: Mass Quadrupole 1}] can cause trouble if the relaxation mesh is not fine enough and uniform'. 

\subsection{Properties of the magneto-thermo-elastic mountains}
\label{SubSec: Mountains on Accreting Neutron Stars}

\begin{figure*}
    \centering
	\includegraphics[width=\textwidth]{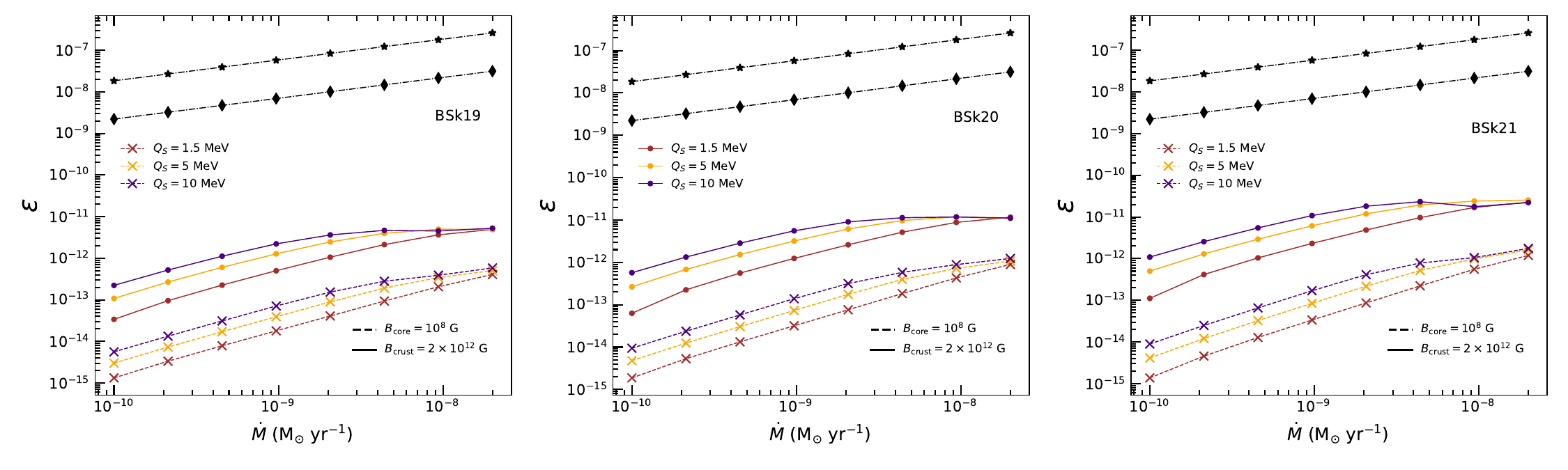}
    \caption[Ellipticity of magnetised neutron stars]{Ellipticity of a number of magnetised neutron stars assuming the BSk19 (\textit{left}), BSk20 (\textit{centre}) and BSk21 (\textit{right}) equations of state (with properties listed in Table \ref{tab:Model properties}) as a function of the mass accretion rate $\dot{M}$. Different amounts of assumed shallow crustal heating, ranging from 1.5 - 10 MeV, are indicated in the legend. Dashed lines (filled circles) denote NS models that assume a $B = 10^8$ G internal \textit{core} toroidal magnetic field, while solid lines (crosses) denote models that assume a $B = 2 \times 10^{12}$ G internal \textit{crustal} toroidal magnetic field. Dashed-dotted lines show the ellipticity required for gravitational-wave torques to determine the spin-equilibrium - via Eq. \eqref{eq: Torque Balance Ellipticity} - of accreting neutron stars with spin frequencies 300 Hz (filled stars) and 700Hz (filled diamonds), as a function of $\dot{M}$.}
    \label{fig:BSk Mass Quadrupoles}
\end{figure*}

With all the necessary pieces assembled, we now discuss the nature of the solutions obtained from the crustal perturbation equations (\ref{eq: z1}) - (\ref{eq: z4}). In Section \ref{Sec: Perturbations in the Thermal Structure} we identified that the assumed level of accretion and shallow heating were crucial to producing large temperature perturbations for a given magnetic field configuration (see Figs \ref{fig:BSk20 Perturbed Temperature Profiles Crust} and \ref{fig:BSk20 Perturbed Temperature Profiles Core}). In Figure \ref{fig:BSk Mass Quadrupoles} we present results for the ellipticity $\varepsilon$ for each of the neutron star models listed in Table \ref{tab:Model properties}, as a function of the mass accretion rate $\dot{M}$, for different assumed values of the shallow crustal heating parameter $Q_{\text{S}}$. Solid lines correspond to models with a $B = 2 \times 10^{12}$ G \textit{crustal} magnetic field, while dashed lines indicate models with a $B = 10^8$ \textit{core} magnetic field.

In general, the deformations generated via the thermal lattice pressure are small ($\varepsilon \lesssim 10^{-11}$) and many orders of magnitude away from the theoretical `maximum' elastic mountain ($\varepsilon_{\text{max}} \sim 10^{-7}$). The smallness of $\varepsilon$, for both the core and crustal magnetic fields (particularly at low accretion rates), can be understood in the following way. First of all, we expect the majority of the mass quadrupole is built in the deep crust, since this is where the majority of the crust's mass is contained (and hence where the largest density perturbations are likely to be generated). In reference to the source term, Equation \eqref{eq: Source Term}, we note that the ratio $P_{\text{th}}/P$ in the deep crust is $\lesssim 10^{-5}$ for the temperatures $ T < 10^{9}$ K relevant to our accreting stars \citep{Jones_2025}. This suggests that the lattice pressure in this region is inconsequential to the total pressure, which is instead overwhelmingly dominated by the contribution from unbound neutrons. 

In the specific case of the core magnetic field, the temperature asymmetry in the deep crust is only $\delta T / T \lesssim 10^{-2} \, \%$ (Fig. \ref{fig:BSk20 Perturbed Temperature Profiles Core}). As discussed in \citet{Hutchins_2023}, we are restricted to magnetic field strengths $B_{\text{core}} \lesssim 10^8$ G in order for the thermal perturbation equations \eqref{eq:Perturbed Temp ODE} - \eqref{eq:Perturbed U ODE} to be valid. While stronger magnetic fields are permitted in the crust, a similar situation ultimately persists: even though temperature asymmetries near the neutron drip point ($\sim 10^{11}$ g cm$^{-3}$) are $\sim 1\%$, in the deep crust it is $\lesssim 0.1\%$ (Fig. \ref{fig:BSk20 Perturbed Temperature Profiles Crust}).

In both cases, the size of the deformation tends to increase with the rate of mass accretion, as well as being enhanced further under greater amounts of shallow crustal heating. This is simply because the amount of heat deposited in the crust is measured per accreted nucleon, and thus the crust is hotter, resulting in larger pressure perturbations. We find the largest ellipticity to be $\varepsilon_{\text{max}} = 2.32\times 10^{-11}$, obtained with the BSk21 equation of state with an accretion rate $\dot{M} = 4.4 \times 10^{-9} \, M_{\odot}$ yr$^{-1}$ and shallow heating term $Q_{\text{S}} = 10$ MeV.

Though, while the general trend is for the ellipticity to increase with the accretion rate, there is in fact a cut off point at $\sim 5 \times 10^{-9} \, M_{\odot}$ yr$^{-1}$, where $\varepsilon$ for different values of the shallow heating parameter converge for a given magnetic field configuration. The reason for this behavior is that at such high accretion rates the amount of heat being deposited into the crust is large enough to melt it at densities $\rho \lesssim 10^{12}$ g cm$^{-3}$, and so only the inner crust contributes to the formation of the mass quadrupole. The background temperature in the deep crust ($\sim 10^{14}$ g cm$^{3}$) is relatively insensitive to the amount of shallow crustal heating, since the additional heat is only deposited at densities $\rho < 10^{10}$ g cm$^{-3}$ \citep{Hutchins_2023}. Since the lattice pressure is very much dependent on the temperature, the ratio $P_{\text{th}} / P$ in the deep crust is also effectively independent of $Q_{\text{S}}$ at high accretion rates. For similar reasons, the temperature perturbations in the deep crust are also largely independent of $Q_{\text{S}}$ (at least at high accretion rates). This is especially true of the perturbations sourced by the crust-only magnetic field, because the inner boundary condition in the perturbed thermal structure model requires that $\delta T \rightarrow 0$ at the crust-core transition as the magnetic field must vanish at the boundary. 

\subsubsection{The torque-balance limit}
\label{SubSubSec: Torque Balance Limit}

In order to gauge the significance of these results, we also plot in each panel of Figure \ref{fig:BSk Mass Quadrupoles} the value of the ellipticity required to reach the gravitational-wave torque-balance limit \eqref{eq: Torque Balance Ellipticity} as a function of the accretion rate, assuming rotation at 300 Hz (filled stars) and 700 Hz (filled diamonds). The ellipticity required to dictate spin-equilibrium via GWs is in the region $\varepsilon_{\text{TB}} \sim 10^{-9} - 10^{-7}$, many orders of magnitude larger than the expected ellipticity generated by the combination of the magnetic field and thermal lattice pressure. 

Although the ellipticity typically increases with the mass accretion rate, the ratio $\varepsilon_{\text{TB}} / \varepsilon $ does not change much with $\dot{M}$. This is because although the hotter crust produces larger deformations, accretion torques on the neutron star also increase ($N_{\text{GW}} \approx \dot{M}\sqrt{GMR}$; \citealt{Ushomirsky_2000}). This necessitates larger GW torques, and therefore an even larger mountain, to maintain torque balance.


The torque-balance limit \eqref{eq: Torque Balance Ellipticity} assumes, however, that 100\% of the spin-down energy from the neutron star is radiated away as gravitational waves. The true picture, however, is complicated by the fact that exactly how matter is transported from an accretion disk to the surface is mediated by the external magnetic field. Under certain circumstances, it is possible for accreted matter to be expelled from the disk rather than being brought down to the surface, thus carrying away angular momentum \citep{Ho_2014}.

The ellipticity for torque balance $\varepsilon_{\text{TB}}$ would therefore necessarily be larger or smaller than Equation \eqref{eq: Torque Balance Ellipticity} depending on whether the external magnetic field is providing spin-up or spin-down torques. Studies such as \citet{Andersson_2005a} indicate the magnetic-spin equilibrium model alone cannot uniquely describe the observed spin-rates of \textit{strongly} accreting neutron stars. It is therefore possible that the mountains generated via the internal magnetic field and the lattice pressure are still playing \textit{some} role in the overall setting of the spin-equilibrium of accreting neutron stars, but are not the dominant source of the spin-down torques. 

With ever-increasing sensitivity of gravitational wave detectors, continuous GW searches are now actually probing the torque balance limit of spinning neutron stars. A recent example of a model-based search for CGWs from the LMXB Scorpius X-1 was performed in \citet{Abbott_2022_Sco-X1}. Whilst no concrete detection was made, upper limits on the gravitational wave strain were set as a function of the GW frequency $f_{\text{GW}} = 2\nu_s$ (since the spin-frequency of Scorpius X-1 is currently unknown). Assuming optimal orientation (i.e.\ where the star's spin axis is orthogonal to the line of sight), the most stringent limit on the GW amplitude was set in the region $f_{\text{GW}} = 100 - 200$ Hz (see their Fig. 6), with $h_0 \sim 4\times 10^{-26}$ (corresponding to an ellipticity $\varepsilon \sim 7 \times 10^{-6}$), which is below the specific torque-balance predictions for Scorpius X-1 (based on measurements of its X-ray flux) from 40 to 200 Hz.  

\subsubsection{Lattice pressure vs. capture layer shifts}
\label{SubSubSec: Crustal lattice pressure vs. capture layer shifts}

In \citet{Jones_2025}, a simple analysis was carried out in order to understand the relative importance of the crustal lattice pressure as compared to shifting capture layers. Order of magnitude estimates of the sizes of the pressure perturbations $\delta P$ were made, for a given (unmodelled) temperature perturbation $\delta T$. It was found that the capture layer pressure perturbations were, when averaged over the full width of the crust, of the order 30 times larger than that of the crustal lattice pressure perturbations. However, this result assumed, as did UCB, the crustal EOS of \citet{Haensel_1990a,Haensel_1990b}, and an average threshold energy of 42 MeV. If instead one were to take into the account of the most recent model of the accreted crust by \citet{Gusakov_2020} (with composition tabulated in \citealt{Potekhin_2023}; see their Tables 1-2), then there would be both fewer capture layers to average over, and their threshold energies would be significantly lower, at around $20$ MeV.

By making use of the formula for the fiducial ellipticity from capture layer shifts in Equation \eqref{eq: UCB Fiducial epsilon}, alongside the results shown in Figure \ref{fig:BSk Mass Quadrupoles}, we may obtain a more robust comparison than was made in \citet{Jones_2025} regarding the importance of the crustal lattice pressure and capture layers shifts for mountain building.

In the capture layer scheme, the total mass quadrupole is a linear sum of the individual quadrupole moments generated in each capture layer \citep{Ushomirsky_2000}. By summing over the total number of capture layers (19) that are listed in Table 2 of \citet{Haensel_1990b}, an upper limit on the total ellipticity from capture layer shifts can be estimated as 

\begin{equation}\label{eq: HZ Capture Layers}
    \Tilde{\varepsilon}_{\text{tot}}^{\, \text{HZ90}} \sim  1.6 \times 10^{-10} \mathlarger{\sum}^{19}_i \bigg[ \frac{E_{\text{cap}}^i}{30 \, \text{MeV}} \biggr]^3 \approx 5.2 \times 10^{-9} \, .
\end{equation}

Depending on the rate of mass accretion and the spin frequency of the star, the above result can be anywhere from $1-4$ orders of magnitude below the torque balance limit. In order to generate ellipticities which \textit{do} probe the torque balance limit, UCB extrapolated to larger values of $E_{\text{cap}}$, and added additional `artificial' capture layers with $A$ and $Z$ $(88, \, 22) \rightarrow (82, \, 20)$ near the bottom of the crust. They found that captures layers with $E_{\text{cap}} \gtrsim 90$ MeV could (even individually) in fact generate mass quadrupoles in excess of $10^{38}$ g cm$^{2}$ ($\varepsilon \sim 10^{-7}$). 

However, in light of recent EOS calculations, if we restrict the calculation above to only sum over capture layers in outer crust (i.e. at densities $ \rho \lesssim 4 \times 10^{11}$ g cm$^{-3}$), then the estimate in Equation \eqref{eq: HZ Capture Layers} reduces to just

\begin{equation}\label{eq: HZ Capture Layers 2}
    \Tilde{\varepsilon}_{\text{tot}}^{\, \text{HZ90}} \lesssim 1.6 \times 10^{-10} \mathlarger{\sum}^{4}_i \bigg[ \frac{E_{\text{cap}}^i}{30 \, \text{MeV}} \biggr]^3 \approx 1.1 \times 10^{-10} \, ,
\end{equation}

\noindent
and is therefore well below even the most conservative estimate of the torque-balance limit.

The `less than' symbol in Equation \eqref{eq: HZ Capture Layers 2} is used to make clear that it is only an upper limit on the ellipticity, even in the outer crust. As described in UCB, so-called ‘sinking penalties’ can arise from the shifting of the capture layers, whereby crustal matter sinks radially rather than spreading out laterally. These penalties can reduce the actual mass quadrupole by as much as a factor of 50 for capture layers in the outer crust (see Fig. 15 of UCB). It is also possible for there to be a change of sign in $Q_{22}$ when going from shallow to deep capture layers (see their Fig. 14), in which case some capture layers in the outer crust may even cancel each other out.

Such effects however can only truly be quantified through the full numerical formulation of the capture layer mechanism, which is beyond the scope of this work. As such we shall continue to make use of Equation \eqref{eq: UCB Fiducial epsilon} in our comparisons. We do note however that a reformulation of UCB's calculations would most certainly be worthwhile in light of the recent equation of state calculations by \citet{Fantina_2018, Fantina_2022} and \citet{Gusakov_2020}, for the reasons just described.

In Table \ref{tab: Ellipticities} we summarize this discussion, and provide estimates of the neutron star ellipticity via capture layer shifts for the cases whereby the capture layers are confined to the outer crust, or extend into the inner crust following the compositional information found in \citet{Haensel_1990a}, \citet{Fantina_2018}, and \citet{Gusakov_2020} (and \citealt{Potekhin_2023}) respectively. The corresponding ellipticity generated via the thermal lattice pressure as presented in this paper for the EOS models of \citet{Fantina_2018} (i.e.\ BSk19, BSk20, and BSk21) are also given. For the purposes of a fair comparison we assume a fixed temperature perturbation $\delta T / T = 1\%$ (rather than sourcing the perturbation via the magnetic field), a time-averaged accretion rate of $10^{-8} M_{\odot}$ yr$^{-1}$, and shallow heating term $Q_{\text{S}} = 0$ MeV, in order to reflect the original choices made in UCB.   

These results indicate that the mass quadrupole generated from the crustal lattice pressure are only comparable in magnitude to that of the displacement of \textit{shallow} capture layers. For a true comparison however, one would ideally compute the mass quadrupole from both the lattice pressure and the displacement of capture layers simultaneously (i.e.\ within the same physical framework), in order to encapsulate the effects of sinking penalties and changes of sign in each individual capture layer. In any case, it is clear that if capture layers \textit{are} present in the inner crust of accreting neutron stars, they would certainly dominate over the lattice pressure.

\begin{table*}
\caption[Ellipticity due to physical capture layer shifts vs. thermal lattice pressures]{Comparison of mountain sizes generated via physical capture layer shifts (as derived by \citealp{Ushomirsky_2000}) and thermal lattice pressures (this work) for three different equation of state models: GC+20 \citep{Gusakov_2020}, HZ+90 \citep{Haensel_1990a, Haensel_1990b}, and F+18 \citep{Fantina_2018, Fantina_2022}. The ellipticity from capture layer shifts is computed - via Eq. \eqref{eq: UCB Fiducial epsilon} - in the full crust and outer crust for the HZ90 and F18 models, and only in the outer crust for the GC20 model. All calculations assume a fixed temperature asymmetry $\delta T / T = 1\%$ and mass accretion rate $\dot{M} = 10^{-8} \, M_{\odot}$ yr$^{-1}$.}
\label{tab: Ellipticities}

\begin{tabular}{@{}ccclc@{}}
\toprule
\toprule
 &
  \multicolumn{2}{c}{\begin{tabular}[c]{@{}c@{}}Ellipticity sourced via capture layer shifts \end{tabular}} &
   &
  \begin{tabular}[c]{@{}c@{}}Ellipticity sourced via thermal lattice pressure \end{tabular} \\ \midrule
EOS          & Full crust            & Outer crust            & \multicolumn{1}{l}{} & Full Crust           \\ \midrule
GC20         & -                     & $2.6 \times 10^{-10}$  &                      & -                    \\
HZ90         & $5.2 \times 10^{-9}$  & $1.1  \times 10^{-10}$ &                      & -                    \\
BSk21 (F18) & $3.9 \times 10^{-9}$  & $9.2 \times 10^{-11}$  &                      &  $2.8\times 10^{-11}$ \\
BSk20 (F18) & $3.1 \times 10^{-9}$  & $9.6 \times 10^{-11}$  &                      & $1.5 \times 10^{-11}$                 \\
BSk19 (F18) & $4.1 \times 10^{-9}$ & $9.2 \times 10^{-11}$  &                      & $8.7\times 10^{-12}$                     \\ \bottomrule \bottomrule
\end{tabular}%

\end{table*}

\subsubsection{Sensitivity of the lattice pressure to crustal composition}
\label{SubSubSec: sensitivity to EoS}

As explained in Section \ref{SubSec: Origin of the mass quadrupole moment} (and \citealt{Jones_2025}), our investigation of crustal lattice pressure as a mechanism for mountain building is motivated in part by the results of \citet{Gusakov_2020}, who predict the absence of inner-crust capture layers. However, all of our numerical results have made use of the EoS calculations of \citet{Fantina_2018}, which do have capture layers in the inner crust.  We used these EoSs as they are well studied, with detailed composition tables being available.  It would be useful to re-run our analysis with an EoS that self-consistently computes the inner crust composition with  the neutron Hydro-static and Diﬀusion (nHD) model of \citet{Gusakov_2020}, so see how the results change.  

In the absence of a detailed composition data for such a model, we can still make a little progress.  As noted by  \citet{Gusakov_2020}, in their analysis, the composition of the inner crust of an accreted neutron star is similar to that of cold-catalysed matter (see their Fig. 2 and surrounding text). This allows us to perform an initial comparison without actually re-creating the nHD model at all, since the BSk pressure-density relations are very similar for both accreted and non-accreted crusts (Fig. \ref{fig:Accreted EOS Analytic Fits}). 

The main compositional difference between an accreted and a non-accreted (catalyzed) inner crust in the BSk model is in the atomic number $Z$. The accreted composition predicts $Z$ steadily decreasing with depth, reaching $Z \sim 14-16$ near the base of the crust; whereas the non-accreted/catalyzed composition keeps a much larger, roughly constant $Z = 40$ through most of the inner crust. The atomic mass, $A$, follows the opposite pattern: $A$ remains modest ($ \lesssim 90$) throughout an accreted crust, while in a catalyzed crust $A$ typically grows substantially larger ($\sim 200$) as density increases. This fact can clearly be seen for each of BSk19, BSk20, and BSk21, by comparing Figs. 2-3 of \citet{Fantina_2018} and Fig. 5 of \citet{Potekhin_2013} for the accreted and non-accreted crusts respectively.

Several other ingredients relevant to the mountain calculation depend on the $(A, Z)$ profile besides the lattice thermal pressure: the shear modulus of the crust scales with the Coulomb coupling parameter of the lattice (Sec. \ref{SubSubSec: Shear Modulus}); and both thermal conductivity and neutrino cooling rates in the crust depend on composition through electron-ion scattering rates, impurity content, and the specific reaction channels available at a given $A$, $Z$, $X_n$ (see Tables 2 and 5 in \citealt{Hutchins_2023}).

We will not attempt to quantify the composition-sensitivity of all these ingredients here, since they depend on multiple competing channels (electron-phonon and electron-impurity scattering, direct/modified Urca-type processes, pair breaking/formation, etc.) whose relative importance shifts with composition in ways that would need dedicated modeling to pin down.  We will content ourselves with a look at the thermal lattice pressure in the non-accreted case.  We can make sue of Equations (14) - (16) in \citet{Potekhin_2013}, which parameterise the values of $Z$, $A$, and $X_n$ in the inner crust for catalyzed matter, as a function of baryon number density.

For the BSk21 equation of state, at a reference density $\rho = 10^{13}$ g cm$^{-3}$, one obtains $(A,Z) = (168,40)$ for the non-accreted crust, and $(A,Z) = (64,16)$ for the accreted crust (Table A.1. in \citealt{Fantina_2018}). For a representative temperature $T = 3\times 10^8$ K, one finds, using our Equations \eqref{eq: pressure perturbation} and \eqref{eq: Lattice Pressure Derivative}, that the fractional changes in the lattice pressure caused by a fractional (Lagrangian) temperature perturbation $\Delta T_{22} / T = 1\%$ are $\sim 1\times 10^{-7}$ and $\sim 3\times 10^{-7}$ for a non-accreted and accreted crust respectively. 

Despite $A$ and $Z$ differing by a factor of $\sim 3$ between the two crusts, the ratio $Z/A$ and $X_{\rm{n}}$ are remarkably similar (0.24 vs. 0.25 and 0.78 vs. 0.74 respectively), and hence so too are the values of $\Delta P_{\rm{th}}/P$; the only other composition-dependent quantity in Equation \eqref{eq: Lattice Pressure Derivative} is the plasma ion frequency, which itself depends on $Z/A$ rather than $Z$ and $A$ separately. Even the pressure-density relations between the two crusts are effectively identical at densities $\rho > 10^{13}$ g cm$^{-3}$, as per Fig. \ref{fig:Accreted EOS Analytic Fits}. 

We take this as support that swapping between traditional and catalyzed-like inner-crust compositions changes thermally-driven quantities in the crust by a similar order-unity factor, rather than by orders of magnitude. Since we have only verified this directly for $\Delta P_{\rm{th}}/P$, a full assessment of magneto-thermo-elastic mountain sizes in the \citet{Gusakov_2020} model would require propagating this composition dependence through the shear modulus, conductivity, neutrino emissivities, etc. as well.

\subsubsection{Thermal mountains vs. magnetic mountains}
\label{SubSubSec: Thermal mountains vs. magnetic mountains}

To put our results further into context, we can also make a comparison to the mountains generated by a ``conventional" \textit{magnetic mountain}; where magnetic forces directly produce the distortion. We divide our discussion into two cases: magnetic fields in the superconducting core, and fields in the (non-superconducting) crust.

For a neutron star with a superconducting core, \citet{Osborne_2020} estimated, from earlier results assembled by \citet{Cutler_2002}, that the ellipticity of a magnetic mountain 

\begin{equation}\label{eq: Magnetic mountain scaling}
    \varepsilon \sim \frac{B_{\rm core} H_{c1} R^4}{GM^2} \approx 2 \times 10^{-9} \, B_{{\rm core},  12} \, ,
\end{equation}

\noindent
where $H_{c1}$ is known as the `first critical field' ($\sim 10^{15}$ G), and $B_{\rm core}$ is the strength of the \emph{core} magnetic field in units of $10^{12}$ G.

Since our results depend on additional parameters other than the magnetic field strength (i.e. the accretion rate, shallow heating, ...), making such a comparison is not straightforward.  For the sake of definiteness, let us consider a uniform internal field strength of $10^{12}$ G, i.e. with a core field of strength $10^{12}$ G, and a crustal toroidal field of the same strength.
In this case, we can compare the magnetic mountain Eq. \eqref{eq: Magnetic mountain scaling} with the thermal mountains in Table \ref{tab: Ellipticities}, sourced by both the thermal lattice pressure, as well as the estimates of $\varepsilon$ from physical capture layer shifts (through Eq. \eqref{eq: UCB Fiducial epsilon}), since $\delta T / T \sim 1\%$ when $\dot{M} \sim \dot{M}_{\text{Edd}}$ and $B_{\text{crust}} = 10^{12}$ G (Fig. \ref{fig:BSk20 Perturbed Temperature Profiles Crust}). 

From Equation \eqref{eq: Magnetic mountain scaling}, one finds $\varepsilon_{\text{mag}} = 2 \times 10^{-9}$ when $B_{\text{crust}} = 10^{12}$ G. Averaging across each EOS model in Table \ref{tab: Ellipticities}, for the "thermal" mountains we have for each individual case:

\begin{itemize}
\renewcommand\labelitemi{--}
  \item $\bar{\varepsilon}_{\text{cap}} \sim 10^{-9}$ for the physical shifts of capture layers in the inner and outer crust (left column),
  
  \item $\bar{\varepsilon}_{\text{cap}} \sim 10^{-10}$ for the physical shifts of capture layers in just the outer crust (centre column),
  
  \item $\bar{\varepsilon}_{\text{lat}} \sim 10^{-11}$ for the thermal lattice pressure in the inner and outer crust  (right column).
  
\end{itemize}

\noindent

Assuming temperature asymmetry at the percent level, this suggests that the magnetic mountains are comparable in size to thermal mountains generated through the displacements of capture layers in the \textit{inner crust} (assuming they exist; recall the discussion in the previous section), but an order of magnitude or so larger if the capture layers are confined to just the outer crust, or when the mass quadrupole is assumed to be built from the thermal crustal lattice pressure.  One must remember, however, that these two sorts of mountain (magnetothermoelastic versus magnetic) are sourced by magnetic fields in different parts of the star (crust versus core), so one cannot definitely conclude that one is stronger than the other unless both field strengths are known.

Note that the results for both magnetic mountains and our thermo-elastic ones are linear in magnetic field strength, so the comparison of relative mountain sizes above is independent of $B$.  However, one cannot scale our numerical results beyond $B_{\rm core} \sim 10^{8}$ G and $B_{\rm crust} \sim 10^{13}$ G, as one would then go out of the domain of validity for our linearisation in the magnetisation parameter $\omega_{B} \tau$; see Section \ref{Sec: Perturbations in the Thermal Structure}.

In the case of magnetic fields in the (non-superconducting) crust, the estimate given by Eq. \eqref{eq: Magnetic mountain scaling} is modified, with the factor $H_{\rm c1} B_{\rm core}$ replaced with $B_{\rm crust}^2$, and one factor of $R$ replaced with $\Delta R$, the crust thickness, to account for the reduction of volume from a sphere of radius $R$ to a shell of thickness $\Delta R$; see e.g. \citet{Cutler_2002}, Eq. (2.3).  We then have

\begin{equation}\label{eq: Magnetic mountain scaling non-super-con}
    \varepsilon \sim \frac{B_{\rm crust}^2  R^3 \Delta R }{GM^2} \approx 2 \times 10^{-13} \, B_{{\rm crust},  12}^2  ,
\end{equation}
where $B_{\rm crust}$ is parameterised in units of $10^{12}$ G.  This is clearly rather small, indicating that magnetic mountains generated within the crust are unlikely to be competitive with the other sorts of mountains discussed above.


\section{Model Improvements}
\label{Sec: Model Improvements}

Having detailed the results of our numerical calculation of thermo-elastic mountain formation, in this section we discuss some possible improvements to the current model. To summarise, our calculation proceeds in essentially three stages: Firstly, the background thermal structure of steadily accreting neutron stars is calculated. We then perturb the star via the addition of a magnetic field to generate a temperature asymmetry, before quantifying how the crust elastically readjusts to the aforementioned asymmetry by coupling to a thermal component of the crustal lattice pressure.

Possible improvements specific to the thermal calculation can be found in \citet{Hutchins_2023}, and so we will not dwell on those in this discussion. Instead, we focus on the elastic and magnetic aspects of our calculation, and identify five areas in which the model could be refined in future work: (i) improved analytical representations of the accreted equation of state, (ii) translation of the current model from Newtonian gravity to full general relativity, (iii) removal of the Cowling Approximation, (iv) extending to a \emph{multifluid} treatment, and (v) a more realistic treatment of the magnetic field.

In the first case, the fundamental building block of our model is the equation of state. While it is the crust of the star that is relevant to building elastic mountains, \citet{Hutchins_2023} demonstrated that the properties of the core can significantly dictate the magnitude of the crustal temperature perturbations which source the (thermal) mountain in the first place. We have chosen to use the BSk family of equations of state since the energy density functionals from which they are derived may be applied not only to nuclear clusters in the crust, but also to homogeneous nuclear matter in the core (see Section 3 of \citet{Fantina_2018} for detail). This has therefore allowed us to describe all regions of the neutron star interior in a consistent manner.

However, as highlighted in Section \ref{SubSec: Analytical Representations of the Accreted Equation of State}, the large density discontinuities present in the heat-producing region $\rho \sim 10^{12} - 10^{13}$ g cm$^{-3}$ initially led to severe issues in achieving numerical convergence when computing the mass quadrupole via Equations \eqref{eq: Mass Quadrupole 1} - \eqref{eq: Mass Quadrupole 2}. One solution to this was to therefore take the approach of approximating the EOS with an analytical fit. Whilst this method was successful in achieving sufficient agreement in Equations \eqref{eq: Mass Quadrupole 1} - \eqref{eq: Mass Quadrupole 2} (to within $\lesssim 1\%$), the procedure produces significant deviations from the `true' pressure-density relation around the site of pycnonuclear reactions due to large density discontinuities. 

The approach of smoothing via a least-squares method, which we adopted, whilst simple, is not necessarily the most robust method to fitting the accreted EOS given the 23-parameter space in Equation \eqref{eq: Analytic EOS Fit} needed to adequately fit the catalysed crust \citep{Potekhin_2013}. It would, therefore, be worthwhile amending the analytical expression (\ref{eq: Analytic EOS Fit}) to further improve the accuracy of the fit, or implement a different method entirely to carry out the optimisation process on the tabulated data. Alternatively, given the implications of the more recent crustal model of \citet{Gusakov_2020}, the ground-state BSk models could serve as an adequate approximation of the inner crust, where the analytical representations are better defined. 

The second point we consider is the fact that currently we use general relativity to compute only the hydrostatic structure of the neutron star. This is required in order to make use of the realistic equations of state described above, which would otherwise have led to unphysical density profiles in Newtonian theory. We do however invoke Newtonian theory to compute both the thermal structure of the star, as well as the elastic readjustment of the crust. Given the compactness of a typical neutron star is of the order $M/R \sim 0.2$, one should expect some fractional errors in the thermal structure when neglecting general relativity. 

We have avoided solving the relativistic (perturbed) Euler equation so as to focus on prescribing a physically motivated mechanism for mountain formation, which has been absent from the literature until now. In principle, this could be rectified by following \citet{Gittins_2021_relativity}, who outline a procedure for obtaining the multipole moments of the neutron star in general relativity. This would also allow for a more consistent treatment of the gravitational acceleration ($g \equiv \nabla_j \Phi$) in our model, where, rather than being defined in terms of an `effective-acceleration' as in Section \ref{Sec: Hydrostatic Structure}, the effects of self-gravity would be encoded in the spacetime metric $g_{ab}$. 

One additional intermediary step, however, could be to remain in the Newtonian setting, but relax the Cowling approximation. This would add an extra term in the Euler equation \eqref{eq: Perturbed Hydrostatic Balance}, which accounts for small perturbations in the gravitational potential, and introduce an additional two ODEs in the system of equations \eqref{eq: z1} - \eqref{eq: z4} that would derive from Poisson's equation for gravity. To elucidate this point further, we note that while not implemented in UCB explicitly, they estimate that inclusion of the effects of self-gravity could decrease the resulting mass quadrupole moment by $20 - 200 \%$. These conclusions were (effectively) reconfirmed in \citet{Haskell_2006} and \citet{Johnson_2013}, with both studies showing that including the perturbations of the gravitational potential reduce the size of the mass quadrupole by a factor of a few\footnote{Note, however, that both these studies are concerned with evaluating the `maximum' mountain that the crust can sustain, rather than exploring the elastic readjustment of the crust from an explicit source term as we have here.}.

One could go beyond our single-fluid treatment and employ a multifluid formalism (see e.g.\ \citet{ac_21}, writing down separate force-balance and continuity equations for the neutron, electron and proton fluids, and for the elastic crust.    The equilibrium solution for such a system would require force balance for each species separately, rather than for the total mass density.  This would include force balance for the free neutrons in the inner crust, the very effect that, in the analysis of \citet{Gusakov_2020}, leads to the elimination of capture layers in the inner crust.  In the absence of such a multifluid treatment, one cannot exclude the possibility that the temperature perturbations in the fluid components lead to a convective motion in the (free) neutrons, i.e.\ one no longer has static solutions, complicating the analysis.

Finally, our treatment of the magnetic field was overly simplistic.  We invoked the finite thermal conductivity to produce our temperature variations, but the electron scattering that this implies will inevitably cause the magnetic field to decay in time.  Focusing on the crust, where the electrical conductivity is lowest, one can estimate the Ohmic diffusion timescale using the standard formula (see e.g. \citet{ss_18}):
\begin{equation}
\label{eq:tau_Ohmic}
    \tau_{\rm Ohmic} = \frac{4 \pi \sigma L^2}{c^2} ,
\end{equation}
where $L$ a typical lengthscale over which the magnetic field varies, and $\sigma$ is the electrical conductivity, given by
\begin{equation}
    \sigma = \frac{n_{\rm e} e^2 \tau}{m_{\rm e}^*} ,
\end{equation}
where $n_{\rm e}$ is the electron number density, $m_{\rm e}^*$ the electron effective mass, and $\tau$ the relaxation time for the electrons.  As summarised in Table 5 of \citet{Hutchins_2023}, in the crust the dominant scattering mechanisms are electron-phonon and electron-impurity, so we have
\begin{equation}
    \tau = \frac{1}{\nu_{\rm ep} + \nu_{\rm eQ}}
\end{equation}
Putting all this together, for a temperature $T \sim 3 \times  10^8$ K, a lengthscale $L \sim 1$ km, and an impurity parameter $Q=1$, and using the parameterised forms of $\nu_{\rm ep}$ and $\nu_{\rm eQ}$ of Table 5 of \citet{Hutchins_2023}, the Ohmic diffusion timescale in the deep crust is $\tau_{\rm Ohmic} \sim 10^6$ years.  

This is much shorter than the $\sim 10^9$ year lifetime of an LMXB.  It follows that some sort of regeneration mechanism process is required to maintain such a crustal field.  However, this is perfectly plausible; LMXBs are believed to be the progenitors of millisecond pulsars, which are observed to have external magnetic fields of order $10^9$ G.  Clearly, some process allows such fields to survive over the LMXB lifetime.  This may be connected with the much higher electrical conductivity in the stellar core, which in turn leads to much longer-lived fields there, and burial of the magnetic field once it has entered the (accreting) LMXB phase.  It may well be that a self-consistent and time-evolving coupling of core and crust is needed to properly model the crustal field, with Ohmic decay, field burial due to accretion, and diffusion of the core field outwards, all contributing to the crustal field strength. A more realistic analysis of thermal mountains would ideally contain such a coupling.

\section{Outlook}
\label{Sec: Summary}

In the Introduction of this paper we outlined two of the primary questions surrounding the feasibility of detecting continuous gravitational-waves from elastic mountains in the near future, namely: the largest possible elastic strains that the crust could feasibly maintain without cracking; and what physical processes inside the star could feasibly lead to the necessary strains forming to begin with.

In answer to the second question, we have sought to produce a self-consistent calculation of the ellipticity of accreting LMXBs due to the formation of what we call `magneto-thermo-elastic' mountains. Many aspects of our calculation have drawn inspiration from the seminal model of thermal mountains presented in \citet{Ushomirsky_2000} (the UCB model), which we have either sought to improve, or re-work entirely. UCB argued that, under certain conditions, it is possible for accreting neutron stars to attain the gravitational torque-balance limit due to shifts of capture layers in regions of the star that are, on average, locally hotter or colder. 

We stress again that these results are predicated on the existence of capture layers well beyond that predicted by modern equations of state. Indeed, when one considers only the capture layers predicted by "realistic" equations of state, the ellipticities can be reduced by as much as three orders of magnitude. It is worth reinforcing that such a reduction also represents an upper limit, since Table \ref{tab: Ellipticities} also assumes a fixed $\delta T / T \sim 1\%$ (i.e. not physically motivated), and the ellipticity from capture layer shifts approximated by Equation \eqref{eq: UCB Fiducial epsilon} scales linearly in $\delta T / T$; nor does Equation \eqref{eq: UCB Fiducial epsilon} take into account the aforementioned 'sinking penalties' and 'canceling out' of shallow capture layers. We have instead explored a method to generate pressure perturbations in the accreted crust of a magnetised neutron star via the crustal lattice pressure.  This has a simpler dependence of temperature on the EOS, and does not require the existence of capture layers at all. Also recall that, while we have not explored it here, the thermal lattice pressure mechanism is also valid for non-accreting neutron stars, unlike the electron capture model (Sec. \ref{SubSubSec: Crustal Lattice Pressure}).   

We have considered the mass quadrupole moment generated in LMXBs which accrete at different rates in order to explore the parameter space derived from observations. In general, we find that even the most optimistic estimates of the ellipticity (i.e.\ in strongly accreting neutron stars) from this mechanism are just $\varepsilon \lesssim 10^{-11}$ (Fig. \ref{fig:BSk Mass Quadrupoles}), and many orders of magnitude away from the torque balance limit, as well as the theoretical upper limit (or `maximum mountain') that currently sits at around $\varepsilon_{\text{max}} \sim 10^{-6} - 10^{-7}$ \citep{Gittins_2021, Gittins_2021_relativity, Morales_2022}. Given that the current observational upper limit on the ellipticity of accreting millisecond pulsars is $\sim 10^{-7}$ \citep{Abbott_2022_AMXP}, it could therefore be the case that detecting `real' thermal mountains will require GW interferometers with much greater sensitivities beyond even near-future capabilities. 

That said, in the continued search for a CGW signal, it may also be beneficial to consider other sources for mountains beyond accreting millisecond pulsars. The GW signal from a rigidly rotating deformed neutron star scales as $h_0 \propto \epsilon \nu_s^2$, and therefore these systems are targeted (at least in part) due to their rapid rotation rates. 

Ultraluminous X-ray sources (ULXs) are a separate class of astrophysical objects (see e.g \citealp{King_2023}), with some thought to contain systems of very strongly accreting neutron stars. Unlike an LMXB, these systems may accrete at super-Eddington rates ($\dot{M} > 2 \times 10^{-8} M_{\odot}$ yr$^{-1}$), and possess magnetic fields in excess of $10^{12}$ G (compared to that of $10^9$ G for a typical LMXB). Not only does this make neutron star ULXs excellent potential sources for \textit{magnetic mountains} (i.e.\ where the mass distortion is supported by Lorentz forces; e.g. \citealp{Cutler_2002}), but specific to the thermo-elastic mountain formation mechanism described here; the temperature asymmetry is linear in magnetic field strength (up to $B \sim 10^{12}$ G; \citealt{Hutchins_2023}), as well as approximately linear in the mass accretion rate. The thermal lattice pressure increases as the temperature increases, and therefore increases for higher accretion rates.  Indeed, a study of magneto-thermo-elastic mountains has in fact been carried out very recently by \citet{Li_2025}, who found potentially detectable levels of gravitational wave emission, if the systems spin sufficiently rapidly.

However, because of their exceedingly strong magnetic fields, the neutron stars in these systems typically spin much more slowly than LMXBs ($\nu_s < 50$ Hz as opposed to $\sim 300 - 700$ Hz). It is therefore not immediately clear whether the greater mass quadrupole generated as a result of strong magnetic fields and enhanced accretion is counteracted by the low spin-rate, and thus not lead to a stronger signal in the detector. If this counteraction is not present, then ULX systems could indeed  be excellent candidates for future targeted searches of advanced detectors such as Einstein telescope \citep{ET_Science_Case_2023} and Cosmic Explorer \citep{Cosmic_Explorer_2023, A_Plus_to_CE_2023}, with their increased sensitivity at low frequency ($\nu_s \sim 50$ Hz $\rightarrow f_{\text{GW}} \sim 100$ Hz).


\section*{Acknowledgements}

TJH acknowledges support from the Science and Technology Facilities Council (STFC) through grant no. ST/T5064121/1. DIJ acknowledges support from the STFC via grant nos. ST/R00045X/1 and APP46132. We also wish to thank the anonymous reviewer, as well as Pep Covas Vidal, for helpful comments on the manuscript and for pointing out typographical errors.

\section*{Data Availability}
The data underlying this article will be shared on reasonable request to the corresponding author.




\bibliographystyle{mnras}
\bibliography{bibliography} 

@INPROCEEDINGS{ss_18,
       author = {{Schmitt}, Andreas and {Shternin}, Peter},
        title = "{Reaction Rates and Transport in Neutron Stars}",
    booktitle = {Astrophysics and Space Science Library},
         year = 2018,
       editor = {{Rezzolla}, Luciano and {Pizzochero}, Pierre and {Jones}, David Ian and {Rea}, Nanda and {Vida{\~n}a}, Isaac},
       series = {Astrophysics and Space Science Library},
       volume = {457},
        month = jan,
        pages = {455},
          doi = {10.1007/978-3-319-97616-7_9},
archivePrefix = {arXiv},
       eprint = {1711.06520},
 primaryClass = {astro-ph.HE},
       adsurl = {https://ui.adsabs.harvard.edu/abs/2018ASSL..457..455S}
}

@ARTICLE{ac_21,
       author = {{Andersson}, Nils and {Comer}, Gregory L.},
        title = "{Relativistic fluid dynamics: physics for many different scales}",
      journal = {Living Reviews in Relativity},
         year = 2021,
        month = dec,
       volume = {24},
       number = {1},
          eid = {3},
        pages = {3},
          doi = {10.1007/s41114-021-00031-6},
archivePrefix = {arXiv},
       eprint = {2008.12069},
 primaryClass = {gr-qc},
       adsurl = {https://ui.adsabs.harvard.edu/abs/2021LRR....24....3A}
}

@article{Abbott_2004,
  title = {Limits on Gravitational-Wave Emission from Selected Pulsars Using LIGO Data},
  author = {Abbott, B. and others},
  collaboration = {LIGO Scientific Collaboration},
  journal = {Phys. Rev. Lett.},
  volume = {94},
  issue = {18},
  pages = {181103},
  numpages = {6},
  year = {2005},
  month = {May},
  publisher = {American Physical Society},
  doi = {10.1103/PhysRevLett.94.181103},
  url = {https://link.aps.org/doi/10.1103/PhysRevLett.94.181103}
}

@article{Abbott_2007,
  title = {Upper limits on gravitational wave emission from 78 radio pulsars},
  author = {Abbott, B. and others},
  collaboration = {LIGO Scientific Collaboration},
  journal = {Phys. Rev. D},
  volume = {76},
  issue = {4},
  pages = {042001},
  numpages = {20},
  year = {2007},
  month = {Aug},
  publisher = {American Physical Society},
  doi = {10.1103/PhysRevD.76.042001},
  url = {https://link.aps.org/doi/10.1103/PhysRevD.76.042001}
}

@article{Abbott_2017_kp,
doi = {10.3847/1538-4357/aa677f},
url = {https://dx.doi.org/10.3847/1538-4357/aa677f},
year = {2017},
month = {apr},
publisher = {The American Astronomical Society},
volume = {839},
number = {1},
pages = {12},
author = {B. P. Abbott and others},
title = {First Search for Gravitational Waves from Known Pulsars with Advanced LIGO},
journal = {\apj}
}

@article{Abbott_2017_kpadd,
  title = {First narrow-band search for continuous gravitational waves from known pulsars in advanced detector data},
  author = {Abbott, B. P. and others},
  collaboration = {LIGO Scientific Collaboration and Virgo Collaboration},
  journal = {Phys. Rev. D},
  volume = {96},
  issue = {12},
  pages = {122006},
  numpages = {20},
  year = {2017},
  month = {Dec},
  publisher = {American Physical Society},
  doi = {10.1103/PhysRevD.96.122006},
  url = {https://link.aps.org/doi/10.1103/PhysRevD.96.122006}
}

@article{Abbott_2017_tensorial_pulsars,
  title = {First Search for Nontensorial Gravitational Waves from Known Pulsars},
  author = {Abbott, B. P. and others},
  collaboration = {LIGO Scientific Collaboration and Virgo Collaboration},
  journal = {Phys. Rev. Lett.},
  volume = {120},
  issue = {3},
  pages = {031104},
  numpages = {13},
  year = {2018},
  month = {Jan},
  publisher = {American Physical Society},
  doi = {10.1103/PhysRevLett.120.031104},
  url = {https://link.aps.org/doi/10.1103/PhysRevLett.120.031104}
}

@article{Abbott_2019_band,
  title = {Narrow-band search for gravitational waves from known pulsars using the second LIGO observing run},
  author = {Abbott, B. P. and others},
  collaboration = {LIGO Scientific Collaboration and Virgo Collaboration},
  journal = {Phys. Rev. D},
  volume = {99},
  issue = {12},
  pages = {122002},
  numpages = {20},
  year = {2019},
  month = {Jun},
  publisher = {American Physical Society},
  doi = {10.1103/PhysRevD.99.122002},
  url = {https://link.aps.org/doi/10.1103/PhysRevD.99.122002}
}

@article{Abbott_2020_Constraint,
doi = {10.3847/2041-8213/abb655},
url = {https://dx.doi.org/10.3847/2041-8213/abb655},
year = {2020},
month = {oct},
publisher = {The American Astronomical Society},
volume = {902},
number = {1},
pages = {L21},
author = {R. Abbott and others},
title = {Gravitational-wave Constraints on the Equatorial Ellipticity of Millisecond Pulsars},
journal = {\apjl},
}

@article{Abac_2026_GWTC4,
doi = {10.3847/2041-8213/ae2c74},
url = {https://doi.org/10.3847/2041-8213/ae2c74},
year = {2026},
month = {jun},
publisher = {The American Astronomical Society},
volume = {1004},
number = {2},
pages = {L22},
author = {A. G. Abac and others},
title = {GWTC-4.0: Updating the Gravitational-wave Transient Catalog with Observations from the First Part of the Fourth LIGO–Virgo–KAGRA Observing Run},
journal = {\apjl},
}

@article{Abbott_2022_narrow_transient,
doi = {10.3847/1538-4357/ac6ad0},
url = {https://dx.doi.org/10.3847/1538-4357/ac6ad0},
year = {2022},
month = {jun},
publisher = {The American Astronomical Society},
volume = {932},
number = {2},
pages = {133},
author = {R. Abbott and others},
title = {Narrowband Searches for Continuous and Long-duration Transient Gravitational Waves from Known Pulsars in the LIGO-Virgo Third Observing Run},
journal = {\apj},}

@ARTICLE{Ghosh_1977,
       author = {{Ghosh}, P. and {Lamb}, F.~K. and {Pethick}, C.~J.},
        title = "{Accretion by rotating magnetic neutron stars. I. Flow of matter inside the magnetosphere and its implications for spin-up and spin-down of the star.}",
      journal = {\apj},
         year = 1977,
        month = oct,
       volume = {217},
        pages = {578-596},
          doi = {10.1086/155606},
       adsurl = {https://ui.adsabs.harvard.edu/abs/1977ApJ...217..578G}
}

@ARTICLE{Ghosh_1979_a,
       author = {{Ghosh}, P. and {Lamb}, F.~K.},
        title = "{Accretion by rotating magnetic neutron stars. II. Radial and vertical structure of the transition zone in disk accretion.}",
      journal = {\apj},
         year = 1979,
        month = aug,
       volume = {232},
        pages = {259-276},
          doi = {10.1086/157285},
       adsurl = {https://ui.adsabs.harvard.edu/abs/1979ApJ...232..259G}
}

@ARTICLE{Ghosh_1979_b,
       author = {{Ghosh}, P. and {Lamb}, F.~K.},
        title = "{Accretion by rotating magnetic neutron stars. III. Accretion torques and period changes in pulsating X-ray sources.}",
      journal = {\apj},
         year = 1979,
        month = nov,
       volume = {234},
        pages = {296-316},
          doi = {10.1086/157498},
       adsurl = {https://ui.adsabs.harvard.edu/abs/1979ApJ...234..296G}
}

@ARTICLE{Ghosh_1978,
       author = {{Ghosh}, P. and {Lamb}, F.~K.},
        title = "{Disk accretion by magnetic neutron stars.}",
      journal = {\apjl},
         year = 1978,
        month = jul,
       volume = {223},
        pages = {L83-L87},
          doi = {10.1086/182734},
       adsurl = {https://ui.adsabs.harvard.edu/abs/1978ApJ...223L..83G}
}

@article{Lattimer_2007,
   title={Neutron star observations: Prognosis for equation of state constraints},
   volume={442},
   ISSN={0370-1573},
   url={http://dx.doi.org/10.1016/j.physrep.2007.02.003},
   DOI={10.1016/j.physrep.2007.02.003},
   number={1-6},
   journal={Physics Reports},
   publisher={Elsevier BV},
   author={Lattimer, J and Prakash, M},
   year={2007},
   month={Apr},
   pages={109–165}
}

@article{Hessels_2006,
   title={A Radio Pulsar Spinning at 716 Hz},
   volume={311},
   ISSN={1095-9203},
   url={http://dx.doi.org/10.1126/science.1123430},
   DOI={10.1126/science.1123430},
   number={5769},
   journal={Science},
   publisher={American Association for the Advancement of Science (AAAS)},
   author={Hessels, J. W. T.},
   year={2006},
   month={Mar},
   pages={1901–1904}
}

@article{Bildsten_1998,
   title={Gravitational Radiation and Rotation of Accreting Neutron Stars},
   volume={501},
   ISSN={0004-637X},
   url={http://dx.doi.org/10.1086/311440},
   DOI={10.1086/311440},
   number={1},
   journal={\apj},
   publisher={American Astronomical Society},
   author={Bildsten, Lars},
   year={1998},
   month={Jul},
   pages={L89–L93}
}

@ARTICLE{Ushomirsky_2000,
       author = {{Ushomirsky}, Greg and {Cutler}, Curt and {Bildsten}, Lars},
        title = "{Deformations of accreting neutron star crusts and gravitational wave emission}",
      journal = {\mnras},
         year = 2000,
        month = dec,
       volume = {319},
       number = {3},
        pages = {902-932},
          doi = {10.1046/j.1365-8711.2000.03938.x},
archivePrefix = {arXiv},
       eprint = {astro-ph/0001136},
 primaryClass = {astro-ph},
       adsurl = {https://ui.adsabs.harvard.edu/abs/2000MNRAS.319..902U}
}

@article{Owen_2005,
   title={Maximum Elastic Deformations of Compact Stars with Exotic Equations of State},
   volume={95},
   ISSN={1079-7114},
   url={http://dx.doi.org/10.1103/PhysRevLett.95.211101},
   DOI={10.1103/physrevlett.95.211101},
   number={21},
   journal={Physical Review Letters},
   publisher={American Physical Society (APS)},
   author={Owen, Benjamin J.},
   year={2005},
   pages={211101},
   month={Nov} }

@ARTICLE{Haskell_2006,
       author = {{Haskell}, B. and {Jones}, D.~I. and {Andersson}, N.},
        title = "{Mountains on neutron stars: accreted versus non-accreted crusts}",
      journal = {\mnras},
         year = 2006,
        month = dec,
       volume = {373},
       number = {4},
        pages = {1423-1439},
          doi = {10.1111/j.1365-2966.2006.10998.x},
archivePrefix = {arXiv},
       eprint = {astro-ph/0609438},
 primaryClass = {astro-ph},
       adsurl = {https://ui.adsabs.harvard.edu/abs/2006MNRAS.373.1423H}
}

@ARTICLE{Johnson_2013,
       author = {{Johnson-McDaniel}, Nathan K. and {Owen}, Benjamin J.},
        title = "{Maximum elastic deformations of relativistic stars}",
      journal = {\prd},
         year = 2013,
        month = aug,
       volume = {88},
       number = {4},
          eid = {044004},
        pages = {044004},
          doi = {10.1103/PhysRevD.88.044004},
archivePrefix = {arXiv},
       eprint = {1208.5227},
 primaryClass = {astro-ph.SR},
       adsurl = {https://ui.adsabs.harvard.edu/abs/2013PhRvD..88d4004J}
}

@ARTICLE{Gittins_2021,
       author = {{Gittins}, F. and {Andersson}, N. and {Jones}, D.~I.},
        title = "{Modelling neutron star mountains}",
      journal = {\mnras},
         year = 2021,
        month = jan,
       volume = {500},
       number = {4},
        pages = {5570-5582},
          doi = {10.1093/mnras/staa3635},
archivePrefix = {arXiv},
       eprint = {2009.12794},
 primaryClass = {astro-ph.HE},
       adsurl = {https://ui.adsabs.harvard.edu/abs/2021MNRAS.500.5570G}
}

@ARTICLE{Gittins_2021_relativity,
       author = {{Gittins}, Fabian and {Andersson}, Nils},
        title = "{Modelling neutron star mountains in relativity}",
      journal = {\mnras},
         year = 2021,
        month = oct,
       volume = {507},
       number = {1},
        pages = {116-128},
          doi = {10.1093/mnras/stab2048},
archivePrefix = {arXiv},
       eprint = {2105.06493},
 primaryClass = {astro-ph.HE},
       adsurl = {https://ui.adsabs.harvard.edu/abs/2021MNRAS.507..116G}
}

@article{Morales_2022,
	doi = {10.1093/mnras/stac3058},
  
	url = {https://doi.org/10.10932Fmnras2Fstac3058},
  
	year = 2022,
	month = {oct},
  
	publisher = {Oxford University Press ({OUP})},
  
	volume = {517},
  
	number = {4},
  
	pages = {5610--5616},
  
	author = {J A Morales and C J Horowitz},
  
	title = {Neutron star crust can support a large ellipticity},
  
	journal = {\mnras}
}

@article{Singh_2020,
	doi = {10.1093/mnras/staa442},
  
	url = {https://doi.org/10.10932Fmnras2Fstaa442},
  
	year = 2020,
	month = {feb},
  
	publisher = {Oxford University Press ({OUP})},
  
	volume = {493},
  
	number = {3},
  
	pages = {3866--3878},
  
	author = {N Singh and B Haskell and D Mukherjee and T Bulik},
  
	title = {Asymmetric accretion and thermal `mountains' in magnetized neutron star crusts},
  
	journal = {\mnras}
}

@article{Osborne_2020,
	doi = {10.1093/mnras/staa858},
  
	url = {https://doi.org/10.1093%2Fmnras%2Fstaa858},
  
	year = 2020,
	month = {mar},
  
	publisher = {Oxford University Press ({OUP})},
  
	volume = {494},
  
	number = {2},
  
	pages = {2839--2850},
  
	author = {E L Osborne and D I Jones},
  
	title = {Gravitational waves from magnetically induced thermal neutron star mountains},
  
	journal = {\mnras}
}

@article{Hutchins_2023,
    author = {Hutchins, T J and Jones, D I},
    title = "{Gravitational radiation from thermal mountains on accreting neutron stars: sources of temperature non-axisymmetry}",
    journal = {\mnras},
    volume = {522},
    number = {1},
    pages = {226-251},
    year = {2023},
    month = {03},
    issn = {0035-8711},
    doi = {10.1093/mnras/stad967},
    url = {https://doi.org/10.1093/mnras/stad967},
    eprint = {https://academic.oup.com/mnras/article-pdf/522/1/226/49883577/stad967.pdf},
}

@article{Jones_2025,
    author = {Jones, D I and Hutchins, T J},
    title = {Crustal lattice pressure as a source of neutron star mountains},
    journal = {Monthly Notices of the Royal Astronomical Society},
    volume = {540},
    number = {3},
    pages = {2349-2358},
    year = {2025},
    month = {05},
    issn = {0035-8711},
    doi = {10.1093/mnras/staf784},
    url = {https://doi.org/10.1093/mnras/staf784},
    eprint = {https://academic.oup.com/mnras/article-pdf/540/3/2349/63212384/staf784.pdf},
}

@article{Cutler_2002,
  title = {Gravitational waves from neutron stars with large toroidal ${B}$ fields},
  author = {Cutler, Curt},
  journal = {\prd},
  volume = {66},
  issue = {8},
  pages = {084025},
  numpages = {6},
  year = {2002},
  month = {Oct},
  publisher = {American Physical Society},
  doi = {10.1103/PhysRevD.66.084025},
  url = {https://doi.org/10.1103\%2Fphysrevd.66.084025}
}

@article{Abbott_2022_AMXP,
  title = {Search for continuous gravitational waves from 20 accreting millisecond x-ray pulsars in O3 LIGO data},
  author = {Abbott, R. and others},
  collaboration = {LIGO Scientific Collaboration, Virgo Collaboration, and KAGRA Collaboration},
  journal = {Phys. Rev. D},
  volume = {105},
  issue = {2},
  pages = {022002},
  numpages = {42},
  year = {2022},
  month = {Jan},
  publisher = {American Physical Society},
  doi = {10.1103/PhysRevD.105.022002},
  url = {https://link.aps.org/doi/10.1103/PhysRevD.105.022002}
}

@article{De_Falco_2017,
	doi = {10.1051/0004-6361/201629575},
  
	url = {https://doi.org/10.10512F0004-63612F201629575},
  
	year = 2017,
	month = {mar},
  
	publisher = {{EDP} Sciences},
  
	volume = {599},
  
	pages = {A88},
  
	author = {V. De Falco and L. Kuiper and E. Bozzo and D. K. Galloway and J. Poutanen and C. Ferrigno and L. Stella and M. Falanga},
  
	title = {The 2015 outburst of the accretion-powered pulsar {IGR} J00291$\mathplus$5934: {INTEGRAL} and$\less$i$\greater$Swift$\less$/i$\greater$observations},
  
	journal = {Astronomy & Astrophysics}
}

@article{Sanna_2016,
	doi = {10.1093/mnras/stw3332},
  
	url = {https://doi.org/10.10932Fmnras2Fstw3332},
  
	year = 2016,
	month = {dec},
  
	publisher = {Oxford University Press ({OUP})},
  
	volume = {466},
  
	number = {3},
  
	pages = {2910--2917},
  
	author = {A. Sanna and F. Pintore and E. Bozzo and C. Ferrigno and A. Papitto and A. Riggio and T. Di Salvo and R. Iaria and A. D{\textquotesingle}A{\`{\i}
} and E. Egron and L. Burderi},
  
	title = {Spectral and timing properties of {IGR} J00291$\mathplus$5934 during its 2015 outburst},
  
	journal = {\mnras}
}

@article{Fantina_2022,
   title={Accreting neutron stars from the nuclear energy-density functional theory. II. Equation of state and global properties},
   url={https://doi.org/10.1051/0004-6361/202243715},
   DOI={10.1051/0004-6361/202243715},
   journal={Astronomy & Astrophysics},
   publisher={EDP Sciences},
   author={Fantina, A. F. and Zdunik, J. L. and Chamel, N. and Pearson, J. M. and Suleiman, S. and Goriely, S.},
   year={2022},
   month={July},
   volume = 665,
	pages = "A74"
}

@ARTICLE{Jaodand_2016,
       author = {{Jaodand}, Amruta and {Archibald}, Anne M. and {Hessels}, Jason W.~T. and {Bogdanov}, Slavko and {D'Angelo}, Caroline R. and {Patruno}, Alessandro and {Bassa}, Cees and {Deller}, Adam T.},
        title = "{Timing Observations of PSR J1023+0038 During a Low-mass X-Ray Binary State}",
      journal = {\apj},
         year = 2016,
        month = oct,
       volume = {830},
       number = {2},
          eid = {122},
        pages = {122},
          doi = {10.3847/0004-637X/830/2/122},
archivePrefix = {arXiv},
       eprint = {1610.01625},
 primaryClass = {astro-ph.HE},
       adsurl = {https://ui.adsabs.harvard.edu/abs/2016ApJ...830..122J}
}

@article{Bildsten_1998_Cumming,
       author = {{Bildsten}, Lars and {Cumming}, Andrew},
        title = "{Hydrogen electron capture in accreting neutron stars and the resulting g-Mode oscillation spectrum}",
      journal = {\apj},
         year = {1998},
        month = {Oct},
       volume = {506},
       number = {2},
        pages = {842-862},
          doi = {10.1086/306279},
archivePrefix = {arXiv},
       eprint = {astro-ph/9807012},
 primaryClass = {astro-ph},
       url = {https://doi.org/10.1086/306279}
}

@ARTICLE{Sato_1979,
       author = {{Sato}, K.},
        title = "{Nuclear Compositions in the Inner Crust of Neutron Stars}",
      journal = {Progress of Theoretical Physics},
         year = 1979,
        month = oct,
       volume = {62},
       number = {4},
        pages = {957-968},
          doi = {10.1143/PTP.62.957},
       adsurl = {https://ui.adsabs.harvard.edu/abs/1979PThPh..62..957S}
}

@article{Schatz_1999,
	doi = {10.1086/307837},
  
	url = {https://doi.org/10.1086%2F307837},
  
	year = 1999,
	month = {oct},
  
	publisher = {American Astronomical Society},
  
	volume = {524},
  
	number = {2},
  
	pages = {1014--1029},
  
	author = {Hendrik Schatz and Lars Bildsten and Andrew Cumming and Michael Wiescher},
  
	title = {The Rapid Proton Process Ashes from Stable Nuclear Burning on an Accreting Neutron Star},
  
	journal = {\apj}
}

@article{Goriely_2010,
  title = {Further explorations of Skyrme-Hartree-Fock-Bogoliubov mass formulas. XII. Stiffness and stability of neutron-star matter},
  author = {Goriely, S. and Chamel, N. and Pearson, J. M.},
  journal = {Phys. Rev. C},
  volume = {82},
  issue = {3},
  pages = {035804},
  numpages = {18},
  year = {2010},
  month = {Sep},
  publisher = {American Physical Society},
  doi = {10.1103/PhysRevC.82.035804},
  url = {https://link.aps.org/doi/10.1103/PhysRevC.82.035804}
}

@ARTICLE{Pearson_2011,
       author = {{Pearson}, J.~M. and {Goriely}, S. and {Chamel}, N.},
        title = "{Properties of the outer crust of neutron stars from Hartree-Fock-Bogoliubov mass models}",
      journal = {\prc},
         year = 2011,
        month = jun,
       volume = {83},
       number = {6},
          eid = {065810},
        pages = {065810},
          doi = {10.1103/PhysRevC.83.065810},
       adsurl = {https://ui.adsabs.harvard.edu/abs/2011PhRvC..83f5810P}
}

@ARTICLE{Pearson_2012,
       author = {{Pearson}, J.~M. and {Chamel}, N. and {Goriely}, S. and {Ducoin}, C.},
        title = "{Inner crust of neutron stars with mass-fitted Skyrme functionals}",
      journal = {\prc},
         year = 2012,
        month = jun,
       volume = {85},
       number = {6},
          eid = {065803},
        pages = {065803},
          doi = {10.1103/PhysRevC.85.065803},
archivePrefix = {arXiv},
       eprint = {1206.0205},
 primaryClass = {nucl-th},
       adsurl = {https://ui.adsabs.harvard.edu/abs/2012PhRvC..85f5803P},
      
}

@ARTICLE{Haensel_1990a,
       author = {{Haensel}, P. and {Zdunik}, J.~L.},
        title = "{Non-equilibrium processes in the crust of an accreting neutron star}",
      journal = {\aap},
         year = 1990,
        month = jan,
       volume = {227},
       number = {2},
        pages = {431-436},
       adsurl = {https://ui.adsabs.harvard.edu/abs/1990A&A...227..431H}
}

@ARTICLE{Haensel_1990b,
       author = {{Haensel}, P. and {Zdunik}, J.~L.},
        title = "{Equation of state and structure of the crust of an accreting neutron star}",
      journal = {\aap},
         year = 1990,
        month = mar,
       volume = {229},
       number = {1},
        pages = {117-122},
       adsurl = {https://ui.adsabs.harvard.edu/abs/1990A&A...229..117H}
}

@article{Fantina_2018,
   title={Crustal heating in accreting neutron stars from the nuclear energy-density functional theory},
   volume={620},
   ISSN={1432-0746},
   url={http://dx.doi.org/10.1051/0004-6361/201833605},
   DOI={10.1051/0004-6361/201833605},
   journal={Astronomy & Astrophysics},
   publisher={EDP Sciences},
   author={Fantina, A. F. and Zdunik, J. L. and Chamel, N. and Pearson, J. M. and Haensel, P. and Goriely, S.},
   year={2018},
   month={Dec},
   pages={A105}
}

@article{2003HZ,
   title={Nuclear composition and heating  in accreting neutron-star crusts},
   volume={404},
   ISSN={1432-0746},
   url={http://dx.doi.org/10.1051/0004-6361:20030708},
   DOI={10.1051/0004-6361:20030708},
   number={2},
   journal={Astronomy & Astrophysics},
   publisher={EDP Sciences},
   author={Haensel, P. and Zdunik, J. L.},
   year={2003},
   month={Jun},
   pages={L33–L36}
}

@article{2008HZ,
   title={Models of crustal heating in accreting neutron stars},
   volume={480},
   ISSN={1432-0746},
   url={http://dx.doi.org/10.1051/0004-6361:20078578},
   DOI={10.1051/0004-6361:20078578},
   number={2},
   journal={Astronomy & Astrophysics},
   publisher={EDP Sciences},
   author={Haensel, P. and Zdunik, J. L.},
   year={2008},
   month={Jan},
   pages={459–464}
}

@article{Fortin_2016,
  title = {Neutron star radii and crusts: Uncertainties and unified equations of state},
  author = {Fortin, M. and Provid\^encia, C. and Raduta, Ad. R. and Gulminelli, F. and Zdunik, J. L. and Haensel, P. and Bejger, M.},
  journal = {Phys. Rev. C},
  volume = {94},
  issue = {3},
  pages = {035804},
  numpages = {21},
  year = {2016},
  month = {Sep},
  publisher = {American Physical Society},
  doi = {10.1103/PhysRevC.94.035804},
  url = {https://link.aps.org/doi/10.1103/PhysRevC.94.035804}
}

@article{Suleiman_2021,
  title = {Influence of the crust on the neutron star macrophysical quantities and universal relations},
  author = {Suleiman, L. and Fortin, M. and Zdunik, J. L. and Haensel, P.},
  journal = {Phys. Rev. C},
  volume = {104},
  issue = {1},
  pages = {015801},
  numpages = {15},
  year = {2021},
  month = {Jul},
  publisher = {American Physical Society},
  doi = {10.1103/PhysRevC.104.015801},
  url = {https://link.aps.org/doi/10.1103/PhysRevC.104.015801}
}

@article{Potekhin_2013,
	author = {{Potekhin, A. Y.} and {Fantina, A. F.} and {Chamel, N.} and {Pearson, J. M.} and {Goriely, S.}},
	title = {Analytical representations of unified equations of state   for neutron-star matter},
	DOI= "10.1051/0004-6361/201321697",
	url= "https://doi.org/10.1051/0004-6361/201321697",
	journal = {A\&A},
	year = 2013,
	volume = 560,
	pages = "A48",
	month = "",
}

@ARTICLE{Gusakov_2020,
       author = {{Gusakov}, M.~E. and {Chugunov}, A.~I.},
        title = "{Thermodynamically Consistent Equation of State for an Accreted Neutron Star Crust}",
      journal = {\prl},
         year = 2020,
        month = may,
       volume = {124},
       number = {19},
          eid = {191101},
        pages = {191101},
          doi = {10.1103/PhysRevLett.124.191101},
archivePrefix = {arXiv},
       eprint = {2004.04195},
 primaryClass = {astro-ph.HE},
       adsurl = {https://ui.adsabs.harvard.edu/abs/2020PhRvL.124s1101G}
}

@article{Zdunik_2008,
	author = {{Zdunik, J. L.} and {Bejger, M.} and {Haensel, P.}},
	title = {Deformation and crustal rigidity
of rotating neutron stars},
	DOI= "10.1051/0004-6361:200810183",
	url= "https://doi.org/10.1051/0004-6361:200810183",
	journal = {A\&A},
	year = 2008,
	volume = 491,
	number = 2,
	pages = "489-498",
}

@article{Ogata_1990,
  title = {First-principles calculations of shear moduli for Monte Carlo--simulated Coulomb solids},
  author = {Ogata, Shuji and Ichimaru, Setsuo},
  journal = {Phys. Rev. A},
  volume = {42},
  issue = {8},
  pages = {4867--4870},
  numpages = {0},
  year = {1990},
  month = {Oct},
  publisher = {American Physical Society},
  doi = {10.1103/PhysRevA.42.4867},
  url = {https://link.aps.org/doi/10.1103/PhysRevA.42.4867}
}

@article{Caplan_2018,
	doi = {10.1103/physrevlett.121.132701},
  
	url = {https://doi.org/10.11032Fphysrevlett.121.132701},
  
	year = 2018,
	month = {sep},
  
	publisher = {American Physical Society ({APS})},
  
	volume = {121},

    pages = {132701},
  
	number = {13},
  
	author = {M.{\hspace{0.167em}
}E. Caplan and A.{\hspace{0.167em}}S. Schneider and C.{\hspace{0.167em}}J. Horowitz},
  
	title = {Elasticity of Nuclear Pasta},
  
	journal = {Physical Review Letters}
}

@article{Deibel_2015,
	doi = {10.1088/2041-8205/809/2/l31},
  
	url = {https://doi.org/10.10882F2041-82052F8092F22Fl31},
  
	year = 2015,
	month = {aug},
  
	publisher = {American Astronomical Society},
  
	volume = {809},
  
	number = {2},
  
	pages = {L31},
  
	author = {Alex Deibel and Andrew Cumming and Edward F. Brown and Dany Page},
  
	title = {A {STRONG} {SHALLOW} {HEAT} {SOURCE} {IN} {THE} {ACCRETING} {NEUTRON} {STAR} {MAXI} J0556-332},
  
	journal = {The Astrophysical Journal}
}

@article{galloway_goodwin_keek_2017, title={Thermonuclear Burst Observations for Model Comparisons: A Reference Sample}, volume={34}, DOI={10.1017/pasa.2017.12}, journal={Publications of the Astronomical Society of Australia}, publisher={Cambridge University Press}, author={Galloway, Duncan K. and Goodwin, Adelle J. and Keek, Laurens}, year={2017}, pages={e019}}

@BOOK{Haensel:2007yy,
       author = {{Haensel}, P. and {Potekhin}, A.~Y. and {Yakovlev}, D.~G.},
        title = "{Neutron Stars 1 : Equation of State and Structure}",
         year = 2007,
       volume = {326},
       adsurl = {https://ui.adsabs.harvard.edu/abs/2007ASSL..326.....H},
      publisher = {Springer},
      series = {Astrophysics and Space Science Library},
}

@ARTICLE{Yakovlev_1980,
       author = {{Yakovlev}, D.~G. and {Urpin}, V.~A.},
        title = "{Thermal and Electrical Conductivity in White Dwarfs and Neutron Stars}",
      journal = {\sovast},
         year = 1980,
        month = jun,
       volume = {24},
        pages = {303},
       adsurl = {https://ui.adsabs.harvard.edu/abs/1980SvA....24..303Y}
}

@article{Radler_2001,
  title = {General-relativistic free decay of magnetic fields in a spherically symmetric body},
  author = {R\"adler, K.-H. and Fuchs, H. and Geppert, U. and Rheinhardt, M. and Zannias, T.},
  journal = {Phys. Rev. D},
  volume = {64},
  issue = {8},
  pages = {083008},
  numpages = {15},
  year = {2001},
  month = {Sep},
  publisher = {American Physical Society},
  doi = {10.1103/PhysRevD.64.083008},
  url = {https://link.aps.org/doi/10.1103/PhysRevD.64.083008}
}

@article{Baiko_2001,
	doi = {10.1103/physreve.64.057402},
  
	url = {https://doi.org/10.1103%2Fphysreve.64.057402},
  
	year = 2001,
	month = {oct},
  
	publisher = {American Physical Society ({APS})},
  
	volume = {64},
  
	number = {5},

    pages = {057402},

  
	author = {D. A. Baiko and A. Y. Potekhin and D. G. Yakovlev},
  
	title = {Thermodynamic functions of harmonic Coulomb crystals},
  
	journal = {Physical Review E}
}

@ARTICLE{Chamel_2008,
  title     = "Physics of neutron star crusts",
  author    = "Chamel, Nicolas and Haensel, Pawel",
  journal   = "Living Rev. Relativ.",
  publisher = "Springer Science and Business Media LLC",
  volume    =  11,
  number    =  1,
  pages     = "10",
  month     =  dec,
  year      =  2008,
  language  = "en"
}

@ARTICLE{Andersson_2005a,
       author = {{Andersson}, N. and {Glampedakis}, K. and {Haskell}, B. and {Watts}, A.~L.},
        title = "{Modelling the spin equilibrium of neutron stars in low-mass X-ray binaries without gravitational radiation}",
      journal = {\mnras},
         year = 2005,
        month = {Aug},
       volume = {361},
       number = {4},
        pages = {1153-1164},
          doi = {10.1111/j.1365-2966.2005.09167.x},
archivePrefix = {arXiv},
       eprint = {astro-ph/0411747},
 primaryClass = {astro-ph},
       url = {https://doi.org/10.1111/j.1365-2966.2005.09167.x}
}

@article{Abbott_2022_Sco-X1,
doi = {10.3847/2041-8213/aca1b0},
url = {https://doi.org/10.3847/2041-8213/aca1b0},
year = {2022},
month = {Dec},
publisher = {The American Astronomical Society},
volume = {941},
number = {2},
pages = {L30},
author = {R. Abbott and others},
title = {Model-based Cross-correlation Search for Gravitational Waves from the Low-mass {X}-Ray Binary {S}corpius {X}-1 in {LIGO O}3 Data},
journal = {The Astrophysical Journal Letters}
}

@misc{Cosmic_Explorer_2023,
      author={Matthew Evans and Alessandra Corsi and Chaitanya Afle and Alena Ananyeva and K. G. Arun and Stefan Ballmer and Ananya Bandopadhyay and Lisa Barsotti and Masha Baryakhtar and Edo Berger and Emanuele Berti and Sylvia Biscoveanu and Ssohrab Borhanian and Floor Broekgaarden and Duncan A. Brown and Craig Cahillane and Lorna Campbell and Hsin-Yu Chen and Kathryne J. Daniel and Arnab Dhani and Jennifer C. Driggers and Anamaria Effler and Robert Eisenstein and Stephen Fairhurst and Jon Feicht and Peter Fritschel and Paul Fulda and Ish Gupta and Evan D. Hall and Giles Hammond and Otto A. Hannuksela and Hannah Hansen and Carl-Johan Haster and Keisi Kacanja and Brittany Kamai and Rahul Kashyap and Joey Shapiro Key and Sanika Khadkikar and Antonios Kontos and Kevin Kuns and Michael Landry and Philippe Landry and Brian Lantz and Tjonnie G. F. Li and Geoffrey Lovelace and Vuk Mandic and Georgia L. Mansell and Denys Martynov and Lee McCuller and Andrew L. Miller and Alexander Harvey Nitz and Benjamin J. Owen and Cristiano Palomba and Jocelyn Read and Hemantakumar Phurailatpam and Sanjay Reddy and Jonathan Richardson and Jameson Rollins and Joseph D. Romano and Bangalore S. Sathyaprakash and Robert Schofield and David H. Shoemaker and Daniel Sigg and Divya Singh and Bram Slagmolen and Piper Sledge and Joshua Smith and Marcelle Soares-Santos and Amber Strunk and Ling Sun and David Tanner and Lieke A. C. van Son and Salvatore Vitale and Benno Willke and Hiro Yamamoto and Michael Zucker},
      year={2023},
      eprint={2306.13745},
      archivePrefix={arXiv},
      primaryClass={astro-ph.IM},
      url={https://arxiv.org/abs/2306.13745}, 
}

@misc{A_Plus_to_CE_2023,
      author={Ish Gupta and Chaitanya Afle and K. G. Arun and Ananya Bandopadhyay and Masha Baryakhtar and Sylvia Biscoveanu and Ssohrab Borhanian and Floor Broekgaarden and Alessandra Corsi and Arnab Dhani and Matthew Evans and Evan D. Hall and Otto A. Hannuksela and Keisi Kacanja and Rahul Kashyap and Sanika Khadkikar and Kevin Kuns and Tjonnie G. F. Li and Andrew L. Miller and Alexander Harvey Nitz and Benjamin J. Owen and Cristiano Palomba and Anthony Pearce and Hemantakumar Phurailatpam and Binod Rajbhandari and Jocelyn Read and Joseph D. Romano and Bangalore S. Sathyaprakash and David H. Shoemaker and Divya Singh and Salvatore Vitale and Lisa Barsotti and Emanuele Berti and Craig Cahillane and Hsin-Yu Chen and Peter Fritschel and Carl-Johan Haster and Philippe Landry and Geoffrey Lovelace and David McClelland and Bram J J Slagmolen and Joshua Smith and Marcelle Soares-Santos and Ling Sun and David Tanner and Hiro Yamamoto and Michael Zucker},
      year={2024},
      eprint={2307.10421},
      archivePrefix={arXiv},
      primaryClass={gr-qc},
      url={https://arxiv.org/abs/2307.10421}, 
}

@article{ET_Science_Case_2023,
doi = {10.1088/1475-7516/2020/03/050},
url = {https://dx.doi.org/10.1088/1475-7516/2020/03/050},
year = {2020},
month = {Mar},
publisher = {},
volume = {2020},
number = {03},
pages = {050},
author = {Michele Maggiore and others},
title = {Science case for the {E}instein {T}elescope},
journal = {Journal of Cosmology and Astroparticle Physics},

}

@BOOK{st_83,
       author = {{Shapiro}, Stuart L. and {Teukolsky}, Saul A.},
       publisher = {John Wiley \& Sons, Ltd},
        title = "{Black holes, white dwarfs and neutron stars. The physics of compact objects}",
         year = 1983,
          doi = {10.1002/9783527617661},
       adsurl = {https://ui.adsabs.harvard.edu/abs/1983bhwd.book.....S}
}

@ARTICLE{McDermott_1988,
       author = {{McDermott}, P.~N. and {van Horn}, H.~M. and {Hansen}, C.~J.},
        title = "{Nonradial Oscillations of Neutron Stars}",
      journal = {\apj},
         year = 1988,
        month = {Feb},
       volume = {325},
        pages = {725},
          doi = {10.1086/166044},
    url = {https://doi.org/10.1086/166044}
}

@article{Jones_Riles_2025,
doi = {10.1088/1361-6382/ada244},
url = {https://doi.org/10.1088/1361-6382/ada244},
year = {2025},
month = {jan},
publisher = {IOP Publishing},
volume = {42},
number = {3},
pages = {033001},
author = {Jones, D I and Riles, K},
title = {Multimessenger observations and the science enabled: continuous waves and their progenitors, equation of state of dense matter},
journal = {Classical and Quantum Gravity}
}

@article{Ho_2014,
    author = {Ho, Wynn C. G. and Klus, H. and Coe, M. J. and Andersson, Nils},
    title = "{Equilibrium spin pulsars unite neutron star populations}",
    journal = {Monthly Notices of the Royal Astronomical Society},
    volume = {437},
    number = {4},
    pages = {3664-3669},
    year = {2013},
    month = {Dec},
    issn = {0035-8711},
    doi = {10.1093/mnras/stt2193},
    url = {https://doi.org/10.1093/mnras/stt2193},
    eprint = {https://academic.oup.com/mnras/article-pdf/437/4/3664/18500792/stt2193.pdf},
}

@article{Potekhin_2023,
	doi = {10.1093/mnras/stad1309},
  
	url = {https://doi.org/10.1093\%2Fmnras\%2Fstad1309},
  
	year = 2023,
	month = {May},
  
	publisher = {Oxford University Press ({OUP})},
  
	volume = {522},
  
	number = {4},
  
	pages = {4830--4840},
  
	author = {A Y Potekhin and M E Gusakov and A I Chugunov},
  
	title = {Thermal evolution of neutron stars in soft X-ray transients with thermodynamically consistent models of the accreted crust},
  
	journal = {\mnras}
}

@ARTICLE{King_2023,
       author = {{King}, Andrew and {Lasota}, Jean-Pierre and {Middleton}, Matthew},
        title = "{Ultraluminous X-ray sources}",
      journal = {New Astronomy Reviews},
         year = 2023,
        month = {Jun},
       volume = {96},
          eid = {101672},
        pages = {101672},
          doi = {10.1016/j.newar.2022.101672},
archivePrefix = {arXiv},
       eprint = {2302.10605},
 primaryClass = {astro-ph.HE},
       url = {https://doi.org/10.1016/j.newar.2022.101672}
}

@ARTICLE{Li_2025,
       author = {{Li}, Hong-Bo and {Kang}, Yacheng and {Xu}, Ren-Xin},
        title = "{Thermal Deformations in Super-Eddington Magnetized Neutron Stars: Implications for Continuous Gravitational-wave Detectability}",
      journal = {\apj},
         year = 2026,
        month = may,
       volume = {1002},
       number = {1},
          eid = {87},
        pages = {87},
          doi = {10.3847/1538-4357/ae5b66},
archivePrefix = {arXiv},
       eprint = {2510.10443},
 primaryClass = {astro-ph.HE},
       adsurl = {https://ui.adsabs.harvard.edu/abs/2026ApJ..1002...87L}
}

@article{Horowitz_2009,
    author = "Horowitz, C. J. and Kadau, Kai",
    title = "{The Breaking Strain of Neutron Star Crust and Gravitational Waves}",
    eprint = "0904.1986",
    archivePrefix = "arXiv",
    primaryClass = "astro-ph.SR",
    doi = "10.1103/PhysRevLett.102.191102",
    journal = "Phys. Rev. Lett.",
    volume = "102",
    pages = "191102",
    year = "2009"
}


\appendix

\section{Perturbation formalism of the crustal displacements} \label{Sec: Perturbation formalism of the crustal displacements}

\subsection{The perturbation equations} \label{Subsec A1: The perturbation equations}

The `unperturbed' background star is assumed to be spherically symmetric. Any perturbed scalar quantity $\Lambda$ (e.g. temperature), either Eulerian  or Lagrangian, may be decomposed into the familiar spherical harmonics. These are written as 

\begin{equation}\label{eq:deltaTSH Eulerrian}
    \delta \Lambda = \sum_{\ell \,=\, 0}^{\infty} \sum_{m \,=\, -\ell}^{\ell} \delta \Lambda_{\ell m}(r) \, Y_{\ell m} (\theta, \phi) \, ,
    \vspace*{3mm}
\end{equation}

\begin{equation}\label{eq:deltaTSH Lagrangian}
    \Delta \Lambda = \sum_{\ell \,=\, 0}^{\infty} \sum_{m \,=\, -\ell}^{\ell} \Delta \Lambda_{\ell m}(r) \, Y_{\ell m} (\theta, \phi) \, ,
    \vspace*{3mm}
\end{equation}

\noindent
respectively, and are related as \citep{st_83}

\begin{equation}\label{eq: Lagrangian to Eulerian Perturbation}
    \Delta \Lambda = \delta \Lambda + \xi^i \nabla_i \Lambda \, .   
\end{equation}

In Newtonian gravity, perturbations of a neutron star with an elastic crust can be described by the (modified) Euler equation

\begin{equation}\label{eq: Hydrostatic Balance}
    0 =  \rho \nabla_j \Phi - \nabla^i \tau_{ij} \, ,
\end{equation}

\noindent
that includes $\tau_{ij}$, the stress-energy tensor of the crust, and where $\Phi$ is the gravitational potential. The stress-energy tensor of the crust is 

\begin{equation}\label{eq: Stress-energy Tensor}
    \tau_{ij} = -Pg_{ij} + t_{ij}, \, 
\end{equation}

\noindent
where $g_{ij}$ is the flat 3-metric and $t_{ij}$ is the (trace-free) shear stress tensor of the solid crust defined by

\begin{equation}\label{eq: Shear-stress Tensor}
    t_{ij} = \mu \, \biggl( \nabla_i \xi_j + \nabla_j \xi_i - \frac{2}{3} \,  g_{ij} \nabla^k \xi_k \biggr) \, ,
\end{equation}

\noindent
where $\mu$ is the shear modulus of the crust (as determined by Equation \eqref{eq: Shear Modulus}; Sec. \ref{SubSubSec: Shear Modulus}). 

We also assume the star to be non-rotating, and as such need only consider polar perturbations of the crust. The appropriate static displacement vector in this case is of the form \citep{Ushomirsky_2000} 

\begin{equation}\label{eq: Displacement Vector}
    \xi^i = \xi^r_{\ell m}(r) \,\hat{r}^i \, Y_{\ell m}  +  \xi^{\perp}_{\ell m}(r) \beta^{-1} r \nabla^i Y_{\ell m} \, ,
\end{equation}

\noindent
where $ \xi^r_{\ell m}$ and $\xi^{\perp}_{\ell m}$ are the radial and tangential components of the displacement respectively, and $\beta = \sqrt{\ell(\ell+1)}$. 

As in Section \ref{Sec: Perturbations in the Thermal Structure}, keeping to linear order and treating the displacement vector $\xi^i$ as a first-order quantity allows us to write down the equations which govern the perturbations of the crust as

\begin{equation}\label{eq: Perturbed Hydrostatic Balance}
    0 = \delta \rho g \hat{r}_j + \nabla^i \delta \tau_{ij}  \, ,
\end{equation}

\noindent
where we have neglected perturbations in gravitational potential (the Cowling approximation), $ |\nabla_j \Phi| = g$ is the acceleration due to gravity (which in the following calculation we define via our `effective-acceleration' $g_{\text{eff}}$ in Equation \eqref{eq: Hybrid G}; Sec. \ref{Sec: Hydrostatic Structure}), and $\delta \rho$ is the Eulerian density perturbation given by \citep{Ushomirsky_2000}

\begin{equation}\label{eq: Perturbed Continuity Equation}
    \delta \rho = - \nabla^i (\rho \xi_i) = - \biggl[ \xi_r \rho' + \rho \biggl(\xi_r' + \frac{2}{r} \xi_r - \frac{\beta}{r} \xi_{\perp} \biggr) \biggr]Y_{\ell m} \, ,
\end{equation}

\noindent
where primes denote radial derivatives.

In order to proceed, we next need to define the perturbed form of the stress-energy tensor in \eqref{eq: Stress-energy Tensor}. The form we use was obtained by UCB, and is given by\footnote{Note in \ref{eq: Transerve traction}) we have corrected a typo from UCB for the last term containing $\xi_r$, \textit{cf}. their 39b).}

\begin{equation} \label{eq: Perturbed stress-energy tensor}
\centering
\begin{split}
    \delta \tau_{ij} = & \, g_{ij} Y_{\ell m} \delta \tau_{rr} + e_{ij} \biggl[ 2\mu \biggl( \frac{1}{r} \xi_r - \xi_r' \biggr) \biggr] Y_{\ell m} \\
    & + f_{ij}\delta \tau_{r\perp} + \Lambda_{ij}\frac{2 \mu \beta}{r} \xi_{\perp} \, ,
\end{split}
\end{equation}

\noindent
where

\begin{equation}\label{eq: Radial traction}
   \delta \tau_{rr} = - \delta P + \mu \biggl( \frac{4}{3} \xi_r' -\frac{4}{3r} \xi_r + \frac{2 \beta}{3r} \xi_{\perp} \biggr)  \, ,
\end{equation}

\begin{equation}\label{eq: Transerve traction}
    \delta \tau_{r\perp} = \mu \biggl( \xi_{\perp}' - \frac{1}{r} \xi_{\perp} + \frac{\beta}{r} \xi_r \biggr) \, ,
\end{equation}

\begin{equation}\label{eq: e_ij}
    e_{ij} = g_{ij} - \hat{r}_i \hat{r}_j \, ,
\end{equation}

\begin{equation}\label{eq: f_ij}
    f_{ij} = \beta^{-1} r \, (\hat{r}_i \, \nabla_j Y_{\ell m} + \hat{r}_j \, \nabla_i Y_{\ell m})  \, ,
\end{equation}

\begin{equation}\label{eq: Lambda_ij}
    \Lambda_{ij} = \beta^{-2} r^2 \, \nabla_i \nabla_j Y_{\ell m} + \beta^{-1}f_{ij}  \, .
\end{equation}

Breaking down the perturbed Euler equation (\ref{eq: Perturbed Hydrostatic Balance}) along $\hat{r}^j$ (i.e.\ in the radial direction) and $\nabla^j Y_{\ell m}$ (the transverse direction) yields the following two expressions

\begin{equation}\label{eq: Radial Piece of Static Balance}
    \delta \rho(r) g =  \delta \tau_{rr}' - \frac{4 \mu}{r} \biggl( \frac{1}{r} \xi_r - \xi_r' \biggr) - \frac{\beta}{r} \delta \tau_{r\perp} + \frac{2 \mu \beta}{r^2} \xi_{\perp}  \, ,
\end{equation}

\noindent
and

\begin{equation} \label{eq: Tangential Piece of Static Balance}
\begin{split}
     0 = \, & \, \delta \tau_{rr} \, + \, 2\mu \biggl( \frac{1}{r} \xi_r - \xi_r' \biggr) \, \\
     & + \, \frac{1}{\beta} \biggl( 3\delta \tau_{r\perp} + r \delta \tau_{r\perp}' \biggr) \, + \, \frac{2\mu \beta}{r} \biggl( \frac{1}{\beta^2} - 1 \biggr) \xi_{\perp} \, ,
\end{split}
\end{equation}

\noindent
from which a set of coupled second-order ODEs for the components of the displacement vector $\xi_r''$ and $\xi_{\perp}''$ may then be obtained using Equations \eqref{eq: Radial traction} and \eqref{eq: Transerve traction} respectively, together with the perturbed continuity equation \eqref{eq: Perturbed Continuity Equation}.

These ODEs may be put into a more suitable form for numerical integration (as well as making the application of boundary conditions more straightforward) by making the following substitutions \citep{McDermott_1988, Ushomirsky_2000}

\begin{gather}\label{eq: Z_i variables}
    \begin{aligned}
    z_1 &= \frac{1}{r} \xi_r \, ,
    & z_2 &= \frac{\Delta \tau_{rr}}{P} = \frac{\delta \tau_{rr}}{P} - z_1 \frac{d \, \text{ln}  \,P}{d \, \text{ln}  \,r} \, ,
    \\
    z_3 &= \frac{1}{\beta r} \xi_{\perp} \, , 
    & z_4 &= \frac{\Delta \tau_{r\perp}}{\beta P} = \frac{\delta \tau_{r\perp}}{\beta P} \, .
    \end{aligned}
\end{gather}

\noindent
This allows us to recast the second-order ODEs in $\xi_r$ and $\xi_{\perp}$ in terms of a set of four coupled \textit{first-order} ODEs for the variables $z_{1-4}$, given as (\textit{cf}. Eqs (43a) - (43d) of UCB)

\begin{subequations}
\label{eq: Perturbation Equations}
 \begin{align}
 \centering
    \frac{d z_1}{d \, \text{ln} \, r} = & \, - \biggl( 1 + \frac{2 \alpha_2}{\alpha_3} \biggr) z_1 + \frac{1}{\alpha_3} \bigl( z_2 + \boldsymbol{\Delta S} \bigr) \nonumber \\
     & \, + \frac{ \ell(\ell+1) \alpha_2}{\alpha_3} z_3 \, ,\label{eq: z1} \vspace*{1in}\\
  \frac{d  z_2}{d  \, \text{ln}  \, r} = & \, \biggl( \tilde{U}\tilde{V} - 4\tilde{V} +  \frac{12 \Gamma \alpha_1}{\alpha_3} \biggr) z_1 \nonumber \\
      & \, + \biggl( \tilde{V} - \frac{4 \alpha_1}{\alpha_3} \biggr) z_2 + \ell(\ell + 1) z_4  - \frac{4\alpha_1}{\alpha_3} \boldsymbol{\Delta S} \nonumber \\
      & \, +  \biggl( \ell(\ell+1)\tilde{V} -  \frac{6 \ell(\ell+1) \Gamma \alpha_1}{\alpha_3} \biggr) z_3 \label{eq: z2} \vspace*{1in}\\
  \frac{d  z_3}{d  \, \text{ln}  \, r} = &  \frac{1}{\alpha_1} z_4 - z_1  \, ,   \label{eq: z3} \vspace*{1in} \\
  \frac{d  z_4}{d  \, \text{ln}  \, r}  = & \, \biggl( \tilde{V} - \frac{6 \Gamma \alpha_1}{\alpha_3} \biggr) z_1 - \frac{\alpha_2}{\alpha_3}z_2 + (\tilde{V}-3)z_4 + \frac{2 \alpha_1}{\alpha_3} \boldsymbol{\Delta S} \nonumber  \\
     &  + \frac{2}{\alpha_3} \bigl\{ [2\ell(\ell+1) - 1]\alpha_1\alpha_2 + 2[\ell(\ell+1)-1]\alpha_1^2 \bigr\} z_3 \, , \label{eq: z4} 
 \end{align}
\end{subequations}

\noindent
where the quantities $\tilde{V}$ and $\tilde{U}$ are defined as

\begin{align} \label{eq: U and V}
    \centering
        \tilde{V} \equiv \frac{\rho g r}{p} = -\frac{d   \, \text{ln}  \,P}{d   \, \text{ln}  \, r} \, , \qquad \tilde{U} \equiv \frac{d \, \text{ln}  \, g}{d \, \text{ln} \, r}  \, + \, 2 \, ,
\end{align}

\noindent
which follow from equations \eqref{eq: TOV equation} and \eqref{eq: derivative Hybrid G} respectively (Sec. \ref{Sec: Hydrostatic Structure}), along with the coefficients $\alpha_{1-3}$ 

\begin{align}
    \centering
        \alpha_1 \equiv \frac{\mu}{P} \, , \qquad \alpha_2 \equiv \Gamma - \frac{2\alpha_1}{3} \, , \qquad \alpha_3 \equiv \Gamma + \frac{4\alpha_1}{3} \, ,
\end{align}

\noindent
where we use the form for the adiabatic index $\Gamma$ as in Section \ref{SubSubSec: Adiabatic Index} and $\mu$ as in Section \ref{SubSubSec: Shear Modulus}.

\noindent

The ODEs \eqref{eq: z1} - \eqref{eq: z4} describe the readjustment of the neutron star crust in response to an inhomogeneous source term (highlighted in bold) 

\begin{equation}\label{eq: Source Term}
    \boldsymbol{\Delta S} = \frac{\partial \, \text{ln}  \,P}{\partial \, \text{ln}  \,T} \biggl|_{\rho} \frac{\Delta T}{T} \, ,
\end{equation}

\noindent
which arises from the temperature dependence of the equation of state in Equation \eqref{eq: Lagrangian pressure}. In our specific case, this source term is derived from the thermal pressure perturbations generated in the crystal lattice from temperature perturbations produced as a result of anisotropic heat conduction as determined by Equations \eqref{eq:Perturbed Temp ODE} - \eqref{eq:V_lm}.

\subsubsection{Eulerian vs Lagrangian temperature perturbations}\label{Subsec A1.1: Boundary Conditions}

Due to our identification of temperature perturbations as being Lagrangian (rather than Eulerian) perturbations of the elastically deformed star (Sec. \ref{Sec: Quadrupolar Deformations of Accreting Neutron Stars}), the system of equations derived by UCB for the case of a $\delta T$ source term include an additional term which does not appear in our system of equations \eqref{eq: z1} - \eqref{eq: z4}. This term is labeled $\alpha_4$ in UCB, and is given by

\begin{equation}\label{eq: alpha_4}
    \alpha_4 = \frac{\partial \, \text{ln}  \,P}{\partial \, \text{ln}  \,T} \biggl|_{\rho} \frac{d   \, \text{ln}  \,T}{d   \, \text{ln}  \,\, r} \, .
\end{equation}

For completeness we show where the additional term $\alpha_4$ originates, and how it compares to our system in reference to the discussion in Section \ref{Sec: Quadrupolar Deformations of Accreting Neutron Stars}. Consider the Lagrangian pressure perturbation equation \eqref{eq: Lagrangian pressure}, for the two cases whereby the temperature perturbations of the fixed crust are identified with the Lagrangian, or Eulerian, temperature perturbations of the deformed crust. These are written as

\begin{equation}
    \Delta P = \frac{\partial P}{\partial \rho} \biggr|_T \, \Delta \rho \, + \,  \frac{\partial P}{\partial T} \biggr|_{\rho} \, \Delta T \, ,
\end{equation}

\noindent
and

\begin{equation}\label{eq: Eulerian Temperature perturbation fixed crust}
    \Delta P = \frac{\partial P}{\partial \rho} \biggr|_T \, \Delta \rho \, + \,  \frac{\partial P}{\partial T} \biggr|_{\rho} \, \biggl[ \delta T + \xi^r_{\ell m} (r) \frac{d T}{d r}  \biggr] 
\end{equation}

\noindent
respectively, where we have made use of the definition in Equation \eqref{eq: Lagrangian to Eulerian Perturbation} to write $\Delta T \equiv \delta T + \xi^r_{\ell m} (r) T'$. These two different identifications of the temperature perturbation may be unified into a single expression for $\Delta P$ by making a change in notation as

\begin{equation}
    \Delta P =  \frac{\partial P}{\partial \rho} \biggr|_T \, \Delta \rho + P \boldsymbol{\Delta S} + Pz_1 \alpha_4 \, , 
\end{equation}

\noindent
where 

\begin{equation}
 \boldsymbol{\Delta S} = \frac{\partial \, \text{ln}  \,P}{\partial \, \text{ln}  \,T} \biggl|_{\rho} \frac{\Delta T}{T} \, , \qquad \alpha_4 = 0
\end{equation}

\noindent
for a Lagrangian temperature perturbation of the elastically deformed star, and 

\begin{equation}
 \boldsymbol{\Delta S} = \frac{\partial \, \text{ln}  \,P}{\partial \, \text{ln}  \,T} \biggl|_{\rho} \frac{\delta T}{T} \, , \qquad  \alpha_4 = \frac{\partial \, \text{ln}  \,P}{\partial \, \text{ln}  \,T} \biggl|_{\rho} \frac{d   \, \text{ln}  \,T}{d   \, \text{ln}  \,\, r} \, ,
\end{equation}

\noindent
for an Eulerian temperature perturbation. In this sense, our perturbation equations are indeed the same as UCB's when $\alpha_4 \equiv 0$, which is the case in their model when the perturbations are due to a
lateral composition gradient $\Delta \mu_{\text{e}}$ - \textit{cf}. their Equation (45).

\subsection{Boundary Conditions} \label{Subsec A2: Boundary Conditions}

The set of equations \eqref{eq: z1} - \eqref{eq: z4} represent a boundary value problem requiring a set of four boundary conditions. Neutron stars are typically composed of a solid crust, bounded by a fluid core and fluid ocean. The computational domain of the calculation is therefore confined to the solid region, since fluids are incapable of supporting shear stresses (i.e. $\mu = 0$ in the fluid).

To be more precise, the inner and outer boundaries are determined by the location of the crust-core transition and the crust-ocean transition respectively. The location of the crust-core transition is a property of the equation of state, and the transitions for BSk19, BSk20, and BSk21 are given in the final column of Table \ref{tab:Model properties}. 

The location of the crust-ocean transition however, while influenced by the equation of state, is determined primarily by the temperature of the crust. Accretion heating can cause the outer layers of the crust to melt. As a result, depending on the rate of accretion of matter onto the neutron star, one should expect that the crust solidifies at different depths (recall Figure \ref{fig:BSk20 Background Temperature Profiles} where the solid regions of the star are denoted by dashed lines). As per Equation \eqref{eq:Coulomb to Thermal}, we define the point at which the crust solidifies as the point whereby the ratio of Coulomb energy to thermal energy exceeds the canonical value 175 for a one component plasma \citep{Haensel:2007yy}. 

The boundary conditions which we shall adopt at these two fluid-solid interfaces are taken from UCB: consider the perturbed Euler equation (\ref{eq: Perturbed Hydrostatic Balance}), and recall the definitions of the substitution values $z_1 - z_4$ given in \eqref{eq: Z_i variables}. At an interface, the radial displacement $\xi_r$ (i.e. $z_1$), as well as both the radial and tangential components of the perturbed traction $\Delta \tau_{rr}$ ($z_2$) and $\Delta \tau_{r\perp}$ ($z_4$) must be continuous. 

At the fluid side of the interface, we require that the Eulerian pressure perturbation $\delta P \equiv 0$. This is a result that we must impose, since we seek to only compute static (i.e. $l \neq 0$) perturbations of the star. By making the Cowling approximation, we effectively ignore perturbations in any fluid region of the star. 
Non-zero pressure perturbations in any fluid regions, in the absence of any perturbations in the gravitational potential, or shear stresses (recall $\mu = 0$ in the fluid) to counterbalance $\delta P$, would lead to displacements of the fluid and no longer give a static solution. 

As a result, at each fluid-solid boundary we must have $\Delta \tau_{rr}(\text{solid}) = \Delta \tau_{rr}(\text{fluid})$. Using the definitions in Equations (\ref{eq: Lagrangian to Eulerian Perturbation}) and (\ref{eq: Radial traction}), and the fact that $\delta P \equiv 0$ on the fluid side of the interface, we can write

\begin{equation} \label{eq: Boundary Condition 1}
\begin{split}
    \qquad & \Delta \tau_{rr}(\text{solid}) = - \, \Delta P(\text{liquid}) \\ & 
     \mspace{-46mu} \implies \delta \tau_{rr}(\text{solid}) - \frac{dP}{dr} \, \xi_{rE} = -\delta P - \frac{dP}{dr} \, \xi_{rF} \\ & 
     \mspace{-46mu} \implies \delta \tau_{rr} \equiv 0 \, . 
\end{split}
\end{equation}

\noindent
where the subscripts $E$ and $F$ denote the radial displacement on the elastic and fluid sides of the interface, respectively. The above result shows that in order to obtain a static solution, in the Cowling approximation, the radial component of the perturbed traction vector $\delta \tau_{rr}$ must be also zero at the top and bottom of the crust. This result, combined with the fact that the tangential component of the perturbed traction vector $\delta \tau_{r\perp}$ must vanish in the fluid (see Eq. (\ref{eq: Transerve traction}) with $\mu = 0$), leads to the set of four boundary conditions \citep{Ushomirsky_2000}

\begin{gather}\label{eq: Boundary Z_i variables}
    \begin{aligned}
    z_2^{\text{crust-ocean}} & = - z_1^{\text{crust-ocean}} \, \frac{d \, \text{ln}  \,P}{d \, \text{ln}  \,r}  \, ,
    & z_4^{\text{crust-ocean}} & = 0 \, ,
    \\
    z_2^{\text{crust-core}} &= - z_1^{\text{crust-core}} \, \frac{d \, \text{ln}  \,P}{d \, \text{ln}  \,r} \, , 
    & z_4^{\text{crust-core}} &= 0 \, .
    \end{aligned}
\end{gather}

\subsection{Obtaining the Ellipticity \label{Subsec A3: Obtaining the quadrupole moment}}

UCB describe two methods of calculating the mass quadrupole moment, which we now briefly describe.   One can use either the perturbed continuity equation (\ref{eq: Perturbed Continuity Equation}), or the radial projection of the perturbed Euler equation in (\ref{eq: Radial Piece of Static Balance}), and writing them down in terms of the substitution variables $z_{1-4}$. These methods yield the respective results

\begin{equation}\label{eq: Density perturbation 1}
    \delta \rho_{\ell m} (r) = -\biggl[z_1 r \frac{d \rho}{dr} \, + \,  \rho \biggl( r\frac{d z_1}{dr} + 3z_1 - \beta^2z_3 \biggr) \biggr] \, ,
\end{equation}

\noindent
or

\begin{equation}\label{eq: Density perturbation 2}
\begin{split}
    \delta \rho_{\ell m} (r) = \frac{1}{g} \biggl[ \frac{d( P[z_2 - \tilde{V} z_1])}{dr} & \, + \, 4\mu\frac{d z_1}{dr} \\ & - \frac{\beta^2}{r} \biggl(P z_4 \, - \, 2\mu z_3 \biggr)\biggr] \, .
\end{split}
\end{equation}

One may then calculate $Q_{22}$ by integrating either Equation \eqref{eq: Density perturbation 1} or Equation \eqref{eq: Density perturbation 2} as per Equation \eqref{eq: Mass Quadrupole}. The resulting expressions are \citep{Ushomirsky_2000}

\begin{equation}\label{eq: Mass Quadrupole 1}
    Q_{22} = 2 \int^{r_{\text{crust-ocean}}}_{r_{\text{crust-core}}} \rho [z_1 + 3z_3] r^4 dr -  (\rho z_1 r^5)|^{r_{\text{crust-ocean}}}_{r_{\text{crust-core}}} \, ,
\end{equation}

\noindent
and

\begin{equation}\label{eq: Mass Quadrupole 2}
\begin{split}
    Q_{22} = - \int^{r_{\text{crust-ocean}}}_{r_{\text{crust-core}}} \frac{\rho}{\Tilde{V}} \biggl\{& 6z_4 - 2\frac{\mu}{p} \biggl[ 2 \frac{d z_1}{d \, \text{ln} \, r} + 6z_3 \biggr] \\ & + (6 - \Tilde{U})(z_2 - \Tilde{V}z_1) \biggr\} r^4 dr \, .
\end{split}
\end{equation}

\noindent
Finally, the ellipticity $\varepsilon$ is then obtained through substitution of either \eqref{eq: Mass Quadrupole 1} or \eqref{eq: Mass Quadrupole 2} into Equation \eqref{eq: Ellipticity}. 

As described in Section \ref{subsec: method of solution}, we constructed a third method of computation, in which the mass quadrupole is calculated simultaneously with the density perturbations, rather than as a later integration once the solution has been obtained; see Equation \eqref{eq: Mass quadrupole ODE}. We found this last method to be the easiest to implement numerically, but used all three as a consistency check.


\bsp	
\label{lastpage}
\end{document}